\documentclass[aps,prl,twocolumn,amsmath,10pt,superscriptaddress,floatfix,nofootinbib]{revtex4-1}
\usepackage{mathtools}
\usepackage{xcolor}
\usepackage{graphicx,bm,tabularx,multirow,enumerate}
\usepackage{dcolumn}
\usepackage{bm}
\usepackage{siunitx}
\usepackage{multirow}
\usepackage{mathtools}

\usepackage[colorlinks=true,linktocpage=true,linkcolor=blue,citecolor=blue]{hyperref}

\newcommand{\bk}[1]{\left(#1\right)}
\newcommand{\squarebk}[1]{\left[#1\right]}
\newcommand{\curlbk}[1]{\left\{#1\right\}}

\newcommand{\ket}[1]{\left|#1\right\rangle}

\newcommand{\defeq}{\vcentcolon=}

\newcommand{\diq}{Q_{1}Q_{2}}
\newcommand{\antidiq}{\bar{Q}_{3}\bar{Q}_{4}}
\newcommand{\anticc}{\bar{c}\bar{c}}
\newcommand{\antibb}{\bar{b}\bar{b}}
\newcommand{\antibc}{\bar{b}\bar{c}}
\newcommand{\gev}{\mathrm{GeV}}

\newcommand{\tev}{\mathrm{TeV}}

\newcommand{\scnu}{\affiliation{{Guangdong Provincial Key Laboratory of Nuclear Science,}\\ Institute of Quantum Matter, South China Normal University, Guangzhou 510006, China}}
\newcommand{\snsc}{\affiliation{Guangdong-Hong Kong Joint Laboratory of Quantum Matter, Southern Nuclear Science Computing Center, South China Normal University, Guangzhou 510006, China}}
\newcommand{\ihep}{\affiliation{Institute of High Energy Physics, Chinese Academy of Sciences, Beijing 100049, China}}

\begin{document}

\title{The lineshape of the compact fully heavy tetraquark}

\author{Zejian Zhuang}\email{zejian.zhuang@m.scnu.edu.cn}
\scnu
\snsc

\author {Ying Zhang}\email{zhangying@m.scnu.edu.cn}
\scnu
\snsc

\author{Yuanzhuo Ma}\email{yuanzhuoma@m.scnu.edu.cn}
\scnu
\snsc

\author{Qian Wang}\email{qianwang@m.scnu.edu.cn, Corresponding author}
\scnu
\snsc
\ihep

\date{\today}

\begin{abstract}
Hadrons and their distributions are the most direct observables 
in experiment, which would shed light on the non-perturbative
mystery of quantum chromodynamics (QCD). As the result, any 
new hadron will challenge our current knowlege on the one hand,
and provide additional inputs on the other hand. The fully heavy
 $cc\bar{c}\bar{c}$ system observed by LHCb recently opens a new 
 era for hadron physics. We first extract the internal structure of the
  fully heavy tetraquarks directly
   from the experimental data, within the compact tetraquark picture. 
   By fitting to the di-$J/\psi$ lineshape, we find that 
   the $X(6900)$ is only cusp effect from the $J/\psi\psi(3770)$ channel.
In addition, there is also a cusp slightly below $6.8~\mathrm{GeV}$
 stemming from the $J/\psi\psi^\prime$ channel. 
The two $0^{++}$ tetraquarks
behave as two resonances above the di-$\eta_c$ and di-$J/\psi$ threshold, respectively.
The $2^{++}$ state is a bound state below the di-$J/\psi$ threshold. 
Furthermore, we find that the $X_{0^{++}}(6035)$ shows 
a significant structure in the di-$\eta_c$ lineshape even 
   after the coupled channel effect. This is an unique feature
   which can distinguish compact $cc\bar{c}\bar{c}$ tetraquark from the loosely hadronic molecules. 
  
\end{abstract}

\maketitle

\section{Introduction}
The formation of hadrons is from
the non-perturbative mechanism of Quantum chromodynamics (QCD),
the theory of strong interaction.
As the result, the properties of hadrons are expected to shed light on
the mystery of non-perturbative dynamics of QCD.
The success of the conventional quark model
attracts the attention of the community to search
for the predicted missing particles for several decades.
The situation breaks up since the observation of the $X(3872)$,
as the first exotic candidate, in 2003. That challenges the conventional quark model
and stimulates research enthusiasm on the so called
exotic hadrons. Up to now, tens of exotic candidates
have been observed by experimental collaborations
(such as LHCb, BESIII, BelleII, JLab, CMS, ATLAS) and numerous studies
~\cite{Chen:2016qju,Chen:2016spr,Dong:2017gaw,Lebed:2016hpi,Guo:2017jvc,Liu:2019zoy,Albuquerque:2018jkn,Yamaguchi:2019vea,Guo:2019twa,Brambilla:2019esw} have been proposed for the understanding of their properties.

Most of them contain a pair of heavy quarks and
locate at heavy quarkonium region. Two competitive scenarios,
i.e. tetraquark and hadronic molecular pictures, are proposed for
their nature based on two clusters, i.e. $cq-\bar{c}\bar{q}^\prime$ and
$c\bar{q}-\bar{c}q^\prime$ respectively. Although great efforts
have been put forward to distinguish these two interpretations for a given particle,
no definite conclusion about any particle yet in the community.
One potential solution is pinning the hope on a fully heavy system,
which is one goal of several experimental collaborations. e.g. the
LHCb~\cite{LHCb:2018uwm,LHCb:2020bwg} and CMS~\cite{CMS:2020qwa} collaborations.
The importance of the fully heavy system is because that
it cannot be classified to several clusters intuitively.
One expects that a comparison of the fully heavy system with
the $c\bar{c}q\bar{q}^\prime$ system would give some hints
for the formation of hadrons.
Luckily, the LHCb collaboration~\cite{LHCb:2020bwg}, recently, reports a narrow structure around $6.9~\gev$
and a broad structure within the range $6.2~\gev\sim 6.8~\gev$ in the
di-$J/\psi$ invariant mass distribution, using the data at $7~\tev$, $8~\tev$ and $13~\tev$ center-of-mass energies.
Because of the observed channel, the quark content of those structures is $cc\bar{c}\bar{c}$.
The narrow structure around $6.9~\gev$ can be explained by a
Breit-Wigner parametrization with and without the interference
between the resonant contribution and the nonresonant contribution~\cite{LHCb:2020bwg}.
This structure is named as the $X(6900)$.

The study of the fully heavy system came back to 1970s which
was motivated by the observation of the $\psi^\prime$~\cite{Iwasaki:1975pv} 
and the severe emerging structures in $e^+e^-$ annihilation~\cite{Chao:1980dv}. 
However, the study becomes a dilemma because of no experimental data. 
The observation of the $X(6900)$~\cite{LHCb:2020bwg} breaks through the situation.
Because of the equal masses of the components, the hadron with $cc\bar{c}\bar{c}$ quarks is
more expected to be a compact one~\cite{Zhao:2020zjh,Faustov:2021hjs,Lu:2020cns,Karliner:2016zzc,Deng:2020iqw,Chen:2016jxd,Zhao:2020nwy,Ke:2021iyh,Zhu:2020xni,Karliner:2020dta,Wang:2019rdo,Weng:2020jao,Berezhnoy:2011xn,Chen:2020lgj,Maciula:2020wri,Wan:2020fsk,Yang:2021hrb,Richard:2021nvn,Szczurek:2021orw,Tiwari:2021tmz,Bedolla:2019zwg,Nefediev:2021pww,Faustov:2020qfm,Wang:2020gmd,Huang:2020dci,Li:2021ygk,Wang:2021kfv,Sonnenschein:2020nwn,Yang:2020atz,Liu:2021rtn,Liu:2019zuc,liu:2020eha}. 
However, none of them fit the di-$J/\psi$ line shape
within this scenario, analogous to that within the molecular picture~\cite{Dong:2020nwy,Liang:2021fzr,Guo:2020pvt,Dong:2021lkh,Wang:2020tpt,Cao:2020gul,Wang:2020wrp,Gong:2020bmg}.
As we know, the experimental events distribution is the most
direct input for theoretical analysis and can provide the underlying structure of 
the interested hadrons, even for the compact object~\cite{Guo:2017jvc,Matuschek:2020gqe}.  

In this work, we first fit the di-$J/\psi$ line shape
with the compact tetraquark picture and extract the corresponding pole positions.
The bare pole positions are extracted from a parameterization
with both chromoelectro and chromomagnetic interactions. 
This study can tell to which extent the compact tetraquark picture can explain the lineshape directly.

\section{Framework}\label{sec:framework}
\subsection{Hamiltonian}
The interaction in the fully heavy tetraquark system can
be described by both chromoelectro and chromomagnetic interaction among
the constituent quarks, i.e.
\begin{equation}\label{eq:hamil}
    H = \sum_{i}\left(m_{i} +T_i\right)+ \sum_{i<j}\squarebk{A_{ij}\bm{\lambda}_{i}\cdot\bm{\lambda}_{j} + \frac{B_{ij}}{m_{i}m_{j}}\bm{\lambda}_{i}\cdot\bm{\lambda}_{j}\bm{S}_{i}\cdot \bm{S}_{j}}
\end{equation}
where $m_{i}$, $\bm{S}_{i}=\frac{1}{2}\bm{\sigma}_i$, $\bm{\lambda}_{i}$ and $T_i$
are mass, spin matrix, Gell-Mann matrix and kinematic energy
for the \emph{i}th quark, respectively. For antiquark, $\bm{\lambda}_{i}$
is replaced by $-\bm{\lambda}_{i}^{*}$.
The expected values of $A_{ij}$ and $B_{ij}$ will be extracted from hadrons.
Due to the non-relativistic property of the fully heavy system,
we take the non-relativistic approximation here,
i.e. neglecting the kinematic terms in Eq.\eqref{eq:hamil}.

\subsection{Wave functions of heavy tetraquark system}
Before proceeding to the solutions of the Hamiltonian of Eq.\eqref{eq:hamil},
one need to analyze the wave function of the fully heavy tetraquark system.
The total wave function of a tetraquark system is constructed by space, flavor, spin and color wave functions individually,
i.e. the total wave function
\begin{equation}\label{eq:total_wave_func}
    \ket{\psi} = \ket{\text{space}}\otimes\ket{\text{flavor}}\otimes\ket{\text{spin}}\otimes\ket{\text{color}}.
\end{equation}
In this work, as we only focus on the ground $S$-wave fully heavy
tetraquarks, the spatial wave function is symmetric and negligible.
Thus, we first construct the spin-color wave function in the diquark-antidiquark configuration.
Here, $\ket{(\diq)_{S_{12}},(\antidiq)_{S_{34}}}_{S}$ are the spin wave functions  with
subscripts $S_{12}$, $S_{34}$ and $S$ the total spins of the first two quarks,
the latter two antiquarks and the sum of them, respectively. The spin wave
functions for various $J^{PC}$s are listed below
\begin{enumerate}[1)]
    \item $J^{PC}=0^{++}:$
          \begin{equation}\label{eq:spin0++}
              \ket{(\diq)_{0},(\antidiq)_{0}}_{0},\;
              \ket{(\diq)_{1},(\antidiq)_{1}}_{0},
          \end{equation}
    \item $J^{PC}=1^{+-}:$
          \begin{equation}\label{eq:spin1+-}
              \begin{aligned}
                   & ~~~~~~ \ket{(\diq)_{1},(\antidiq)_{1}}_{1}                                                            \\
                   & \frac{1}{\sqrt{2}}\squarebk{\ket{(\diq)_{1},(\antidiq)_{0}}_{1}-\ket{(\diq)_{0},(\antidiq)_{1}}_{1}},
              \end{aligned}
          \end{equation}
    \item $J^{PC}=1^{++}:$
          \begin{equation}\label{eq:spin1++}
              \frac{1}{\sqrt{2}}\left(\ket{(\diq)_{1},(\antidiq)_{0}}_{1}+\ket{(\diq)_{0},(\antidiq)_{1}}_{1}\right),
          \end{equation}
    \item $J^{PC}=2^{++}:$
          \begin{equation}\label{eq:spin2++}
              \ket{(\diq)_{1},(\antidiq)_{1}}_{2}.
          \end{equation}
\end{enumerate}
As we only consider $S$-wave fully tetraquark system in
this work, the orbital angular momentum does not appear in the above equations.
The color confinement tells us that all the observed hadrons are color singlet,
which gives the potential color wave functions below
\begin{equation}\label{eq:color66bar}
    \ket{(\diq)^{6},(\antidiq)^{\bar{6}}}^{1},
\end{equation}
\begin{equation}\label{eq:color3bar3}
    \ket{(\diq)^{\bar{3}},(\antidiq)^{3}}^{1}.
\end{equation}
Here the superscripts $6(\bar{6})$ and $\bar{3}(3)$ denote the corresponding irreducible
representations of color SU(3) group for diquark(antidiquark).
Accordingly, $1$ stands for color singlet. In total, the spatial and color wave functions of $S$-wave ground
fully heavy systems are symmetric and antisymmetric,
which leaves the product of spin and flavor wave functions
symmetric due to Pauli principle.
Flavor wave function is symmetric for full-charm(bottom) tetraquarks
and could be symmetric or anti-symmetric for the $bc\bar{b}\bar{c}$ tetraquarks.
As the result, all the potential total wave functions for various $J^{PC}$ are collected in Table.~\ref{tab:wave_func_tetra}.
\begin{table*}[!htbp]
    \caption{\label{tab:wave_func_tetra}The bases of fully heavy tetraquarks, where \{\} and []
        denote the symmetric and antisymmetric flavor functions, respectively, of diquark(antidiquark). The subscripts and superscripts
        are for the irreducible representations in spin and color spaces, respectively.  }
    \begin{tabular}{@{}cccccc@{}}
        \hline\hline
        $J^{PC}$ & Tetraquark                                          & \multicolumn{2}{c}{Wave function}                                                                                                                                                                                                                                                               \\\hline
        \multirow{4}*{$0^{++}$}
                 & $cc\anticc$                                         & $\ket{\curlbk{cc}_{0}^{6}\curlbk{\anticc}_{0}^{\bar{6}}}_{0}^{1}$                                                                          & $\ket{\curlbk{cc}_{1}^{\bar{3}}\curlbk{\anticc}_{1}^{3}}_{0}^{1}$                                                                          & ~ & ~ \\
                 & $bb\antibb$                                         & $\ket{\curlbk{bb}_{0}^{6}\curlbk{\antibb}_{0}^{\bar{6}}}_{0}^{1}$                                                                          & $\ket{\curlbk{bb}_{1}^{\bar{3}}\curlbk{\antibb}_{1}^{3}}_{0}^{1}$                                                                          & ~ & ~ \\
                 & \multirow{2}*{$bc\antibc$}
                 & $\ket{[bc]_{1}^{6}[\antibc]_{1}^{\bar{6}}}_{0}^{1}$ & $\ket{\curlbk{bc}_{0}^{6}\curlbk{\antibc}_{0}^{\bar{6}}}_{0}^{1}$                                                                                                                                                                                                                               \\
        ~        & ~                                                   & $\ket{\curlbk{bc}_{1}^{\bar{3}}\curlbk{\antibc}_{1}^{3}}_{0}^{1}$                                                                          & $\ket{[bc]_{0}^{\bar{3}}[\antibc]_{0}^{3}}_{0}^{1}$
        \\ \hline
        \multirow{4}*{$1^{+-}$}
                 & $cc\anticc$                                         & $\ket{\curlbk{cc}_{1}^{\bar{3}}\curlbk{\anticc}_{1}^{3}}_{1}^{1}$                                                                                                                                                                                                                               \\
                 & $bb\antibb$                                         & $\ket{\curlbk{bb}_{1}^{\bar{3}}\curlbk{\antibb}_{1}^{3}}_{1}^{1}$                                                                                                                                                                                                                               \\
                 & \multirow{2}*{$bc\antibc$}
                 & $\ket{[bc]_{1}^{6}[\antibc]_{1}^{\bar{6}}}_{1}^{1}$ & $\frac{1}{\sqrt{2}}\bk{\ket{[bc]_{1}^{6}\curlbk{\antibc}_{0}^{\bar{6}}}_{1}^{1}-\ket{\curlbk{bc}_{0}^{6}[\antibc]_{1}^{\bar{6}}}_{1}^{1}}$                                                                                                                                                      \\
        ~        & ~                                                   & $\ket{\curlbk{bc}_{1}^{\bar{3}}\curlbk{\antibc}_{1}^{3}}_{1}^{1}$                                                                          & $\frac{1}{\sqrt{2}}\bk{\ket{\curlbk{bc}_{1}^{\bar{3}}[\antibc]_{0}^{3}}_{1}^{1}-\ket{[bc]_{0}^{\bar{3}}\curlbk{\antibc}_{1}^{3}}_{1}^{1}}$
        \\\hline
        $1^{++}$ & $bc\antibc$                                         & $\frac{1}{\sqrt{2}}\bk{\ket{[bc]_{1}^{6}\curlbk{\antibc}_{0}^{\bar{6}}}_{1}^{1}+\ket{\curlbk{bc}_{0}^{6}[\antibc]_{1}^{\bar{6}}}_{1}^{1}}$ & $\frac{1}{\sqrt{2}}\bk{\ket{\curlbk{bc}_{1}^{\bar{3}}[\antibc]_{0}^{3}}_{1}^{1}+\ket{[bc]_{0}^{\bar{3}}\curlbk{\antibc}_{1}^{3}}_{1}^{1}}$
        \\\hline
        \multirow{3}*{$2^{++}$}
                 & $cc\anticc$                                         & $\ket{\curlbk{cc}_{1}^{\bar{3}}\curlbk{\anticc}_{1}^{3}}_{2}^{1}$                                                                                                                                                                                                                               \\
                 & $bb\antibb$                                         & $\ket{\curlbk{bb}_{1}^{\bar{3}}\curlbk{\antibb}_{1}^{3}}_{2}^{1}$                                                                                                                                                                                                                               \\
                 & $bc\antibc$                                         & $\ket{[bc]_{1}^{6}[\antibc]_{1}^{\bar{6}}}_{2}^{1}$                                                                                        & $\ket{\curlbk{bc}_{1}^{\bar{3}}\curlbk{\antibc}_{1}^{3}}_{2}^{1}$
        \\\hline\hline
    \end{tabular}
\end{table*}
\subsection{\label{sec:parameters}Parameters}
\begin{figure}[!htbp]
    \centering
    \includegraphics[width=70mm]{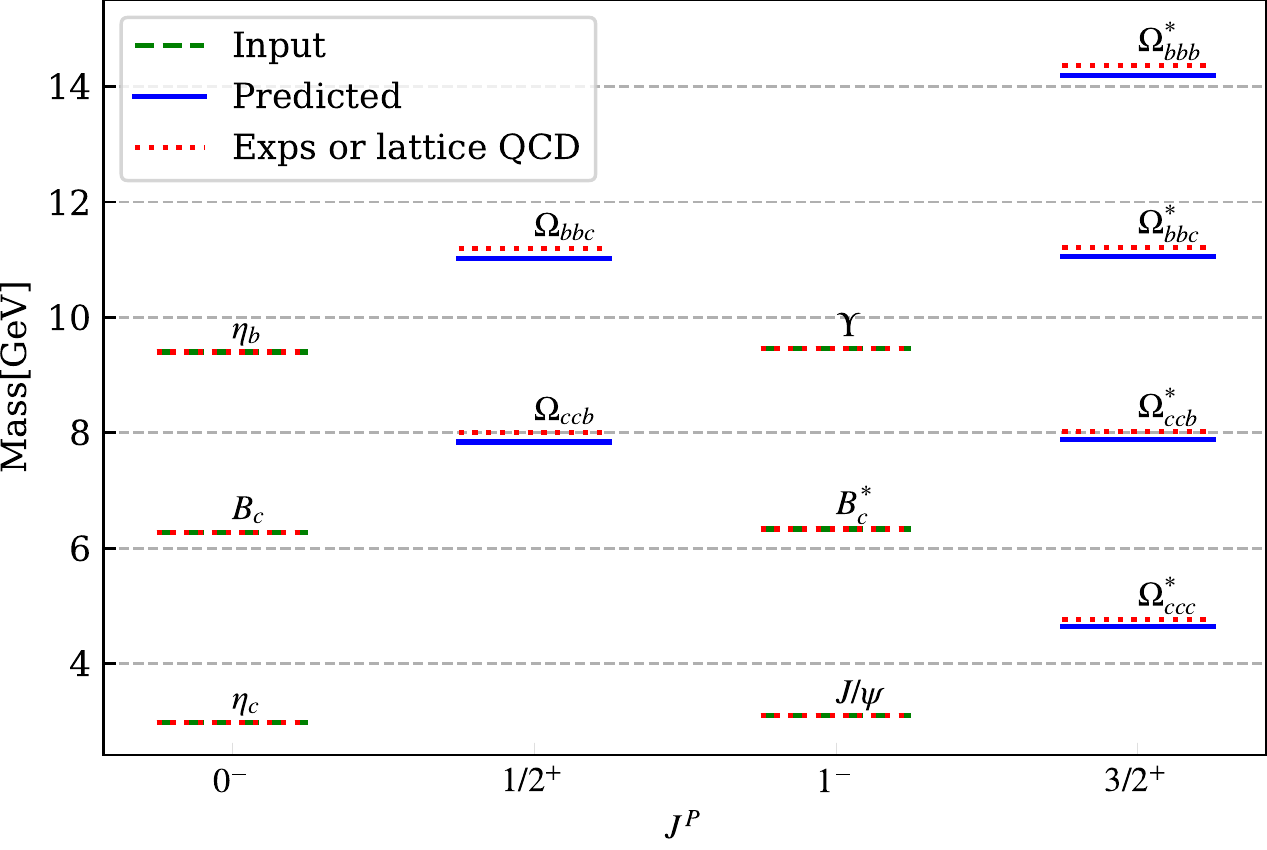}
    \caption{\label{fig:hadron_m}
    The comparison of normal heavy hadrons with
    the predicted values from our framework.
    The red dotted lines are results from either PDG or
    lattice QCD group. The green dashed and blue solid lines are our inputs
    and predictions, respectively. The values of $J/\psi$, $\eta_c$, $\Upsilon$
    and $\eta_b$ masses are from PDG\cite{ParticleDataGroup:2020ssz}. Those
    of $B_c^{(*)}$, $\Omega_{bbc}^{(*)}$ and $\Omega_{ccb}^{(*)}$ are from Ref.~\cite{Mathur:2018epb}
    The masses of $\Omega_{bbb}^*$ and $\Omega_{ccc}^*$ are from Ref.~\cite{Meinel:2010pw} and Ref.~\cite{Padmanath:2013zfa}, respectively.
    }
\end{figure}

In this work, we do not aim at solving Schr\"odinger equation explicitly,
but using a parametrization scheme to extract the mass spectra of
the $S$-wave ground fully heavy tetraquarks. Thus
to investigate the mass spectra,
the expectation values of the parameters $A_{ij}$ and $B_{ij}$
for various systems should be extracted.
As discussed in the above,
the contributions from color and spin spaces have been factorized
out by the $\bm{\lambda}$ and $\bm{S}$ matrixes, respectively. In addition,
the contribution of flavor part can be obtained via the corresponding flavor
wave functions in Table~\ref{tab:wave_func_tetra}. The residue contribution is only from
the spacial part via
\begin{eqnarray}
    \mathcal{A}_{ij}\coloneqq\langle A_{ij}\rangle\\
    \mathcal{B}_{ij}\coloneqq\langle B_{ij}\rangle,
\end{eqnarray}
with $\langle\cdots\rangle$ the expected values of the spacial wave functions.
As we only consider $S$-wave ground tetraquarks, the expected values
can be approximated constants and extracted from $S$-wave ground
pseudoscalar and vector mesons. Due to the symmetry of color and spin spaces,
these expected values have the relation 
\begin{eqnarray}
    \mathcal{A}_{12}&=&\mathcal{A}_{34},\quad \mathcal{A}_{13}=\mathcal{A}_{24}=\mathcal{A}_{14}=\mathcal{A}_{23},\\
    \mathcal{B}_{12}&=&\mathcal{B}_{34},\quad \mathcal{B}_{13}=\mathcal{B}_{24}=\mathcal{B}_{14}=\mathcal{B}_{23}.
\end{eqnarray}
For the parameters $\mathcal{A}_{bc}$ and $\mathcal{B}_{bc}$, the masses of
either $c\bar{b}$ or $b\bar{c}$ mesons are as inputs. However, only one $c\bar{b}$ state, i.e., $B_{c}$
is observed in experiments~\cite{CDF:1998axz,McNeil:1997qa}.
Alternatively, the theoretical results of Ref.~\cite{Ortega:2020uvc} is
used as an input. All the input masses are collected in Table~\ref{tab:hadron_m}.
Here the heavy quark masses $m_{i}$ are fixed to the values
\begin{equation}\label{eq:heavy_quark_m}
    m_{c} = 1.5~\text{GeV},\; m_{b} = 5~\text{GeV},
\end{equation}
in the constituent quark model~\cite{Griffiths:2008zz}. In the end,
\begin{table}[ht]
    \caption{\label{tab:hadron_m} Masses and errors of heavy mesons used for extracting the parameters.
        The values of the first four mesons are extracted from Refs.~\cite{ParticleDataGroup:2020ssz}.}
    \begin{tabular}{@{}ccccccc@{}}
        \hline\hline
        Mass ($\mathrm{MeV}$) & $\eta_{c}$ & $J/\psi$ & $\eta_{b}$ & $\Upsilon$ & $B_{c}$~\cite{Ortega:2020uvc} & $B_{c}^{*}$~\cite{Ortega:2020uvc} \\\hline
        $m$                   & 2983.9     & 3096.9   & 9398.7     & 9460.3     & 6276                                  & 6331                                      \\
        $\Delta m$            & 0.4        & 0.006    & 2          & 0.26       & 7                                     & 7
        \\
        \hline\hline
    \end{tabular}
\end{table}
the mass formulae for extracting parameters are
\begin{align}
    M_{0} & = \sum_{i=c,b} m_{i} + \sum_{i,j=c,b}\squarebk{\frac{-16}{3}\mathcal{A}_{ij}+\frac{4\mathcal{B}_{ij}}{m_{i}m_{j}}},\label{eq:sudoMeson}              \\
    M_{1} & = \sum_{i=c,b} m_{i} + \sum_{i,j=c,b}\squarebk{\frac{-16}{3}\mathcal{A}_{ij}+\frac{-4}{3}\frac{\mathcal{B}_{ij}}{m_{i}m_{j}}},\label{eq:vectorMeson}
\end{align}
where $M_{0}$ and $M_{1}$ are the masses of the pseudoscalar and vector meson,
respectively. With Eqs.(\ref{eq:sudoMeson}-\ref{eq:vectorMeson}), the parameters $\mathcal{A}_{ij}$ and $\mathcal{B}_{ij}$
can be obtained and listed in Table~\ref{tab:paras}.
Before proceeding, we also check the applicability
of our framework by applying to triply heavy baryon systems.
In the same framework, we obtain the mass formulae
\begin{widetext}
    \begin{eqnarray}
        M_{\frac{1}{2}}&=&2 m_{Q}+m_{Q'}-\frac{8}{3}\left(2 \mathcal{A}_{Q Q^{\prime}}+\mathcal{A}_{Q Q}\right)+\left(-\frac{8}{3}\right)\left(\frac{\mathcal{B}_{Q Q}}{4 m_{Q}^{2}}-\frac{\mathcal{B}_{Q Q^{\prime}}}{m_{Q} m_{Q^{\prime}}}\right)\\
        M_{\frac{3}{2}}&=&2 m_{Q}+m_{Q^{\prime}}-\frac{8}{3}\left(2 \mathcal{A}_{Q Q^{\prime}}+\mathcal{A}_{Q Q}\right)+\left(-\frac{8}{3}\right) \frac{1}{4}\left(\frac{\mathcal{B}_{Q Q}}{m_{Q}^{2}}+\frac{2 \mathcal{B}_{Q Q^{\prime}}}{m_{Q} m_{Q}^{\prime}}\right)
    \end{eqnarray}
\end{widetext}
for $J=\frac{1}{2}, \frac{3}{2}$ $QQQ^\prime$ heavy baryons
and
\begin{eqnarray}
    M_{\frac{3}{2}}&=&3 m_{Q}-\frac{8}{3}\left(3 \mathcal{A}_{Q Q}+\frac{3}{4} \frac{\mathcal{B}_{Q Q}}{m_{Q Q}^{2}}\right)
\end{eqnarray}
for $J=\frac{3}{2}$ $QQQ$ heavy baryons.
With the extracted parameters, we obtain
the masses of the $\Omega_{ccb}^{(*)}$, $\Omega_{bbc}^{(*)}$, $\Omega^{*}_{ccc}$ and $\Omega^*_{bbb}$
and compare with the predicted results by lattice QCD
simulation~\cite{Meinel:2010pw,Padmanath:2013zfa,Mathur:2018epb} as shown in Fig.~\ref{fig:hadron_m}.
From Fig.~\ref{fig:hadron_m}, one can see that the masses of triply heavy baryons
are very close to results given by lattice QCD, which indicates the applicability of our framework.
\begin{table*}[!htbp]
    \caption{\label{tab:paras} The values and errors of the parameters $\mathcal{A}_{ij}$ and $\mathcal{B}_{ij}$ in this work. }
    \begin{tabular}{@{}cccccccc@{}}
        \hline\hline
        ~          & \multicolumn{3}{c}{$\mathcal{A}_{QQ^{(')}}$[GeV]} & \multicolumn{3}{c}{$\mathcal{B}_{QQ^{(')}}\text{[GeV]}^{3}$}                                                                                     \\\hline
        Parameters & $\mathcal{A}_{cc}$                                & $\mathcal{A}_{bb}$                                           & $\mathcal{A}_{cb}$ & $\mathcal{B}_{cc}$ & $\mathcal{B}_{bb}$ & $\mathcal{B}_{cb}$ \\\hline
        Value      & -0.01287                                          & 0.10408                                                      & 0.03427            & -0.04767           & -0.28875           & -0.07734           \\
        Error      & \num{2e-5}                                        & \num{e-4}                                                    & \num{1.04e-3}      & \num{2.1e-4}       & \num{9.45e-3}      & \num{1.392e-2}
        \\
        \hline\hline
    \end{tabular}
\end{table*}

\section{\label{sec:results and discuss}Numerical results and discussions}

\subsection{Mass spectra of fully heavy tetraquark system with the non-relativistic parametrization}
\begin{figure}[!htbp]
    \centering
    \includegraphics[width=70mm]{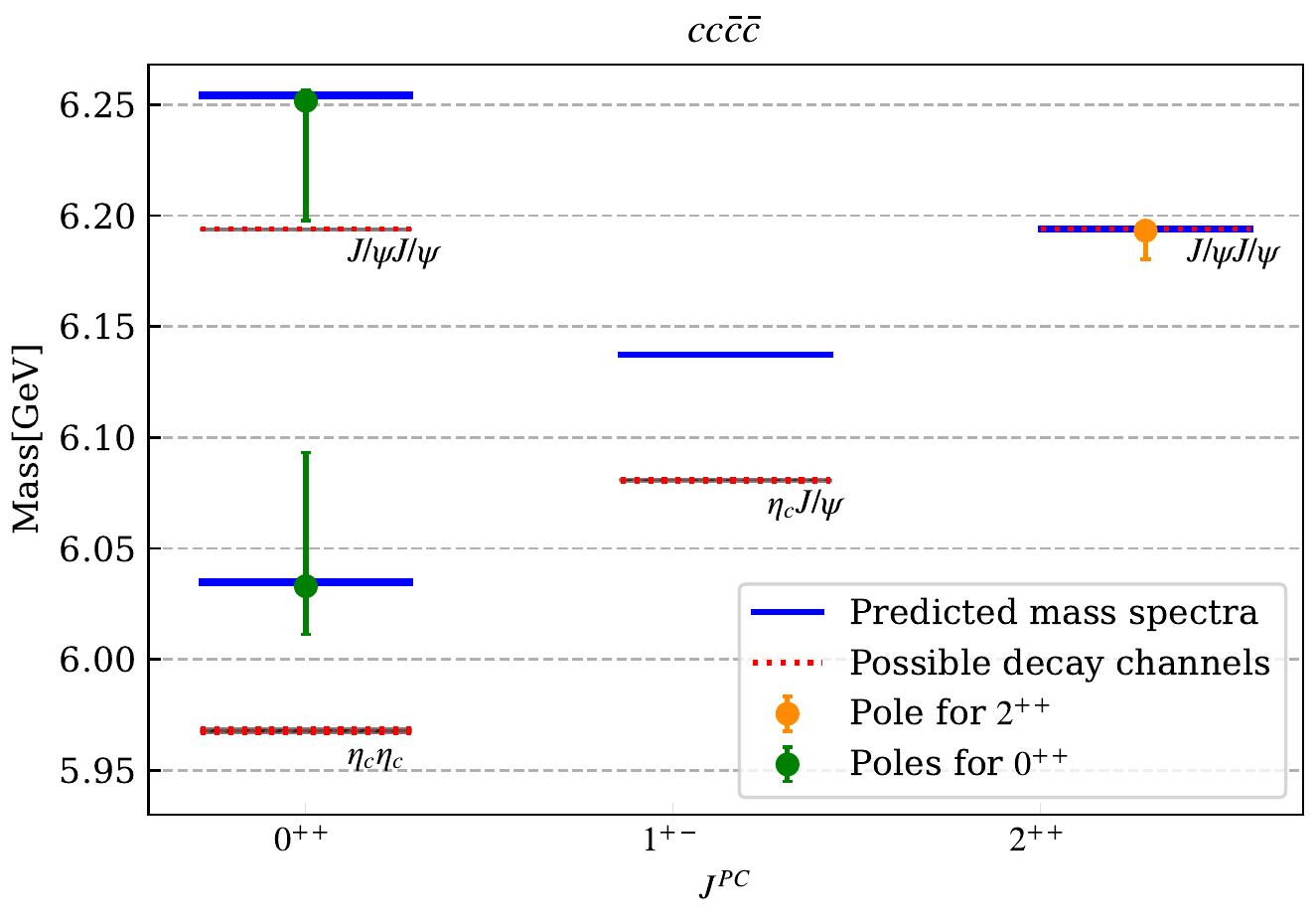}
    \includegraphics[width=70mm]{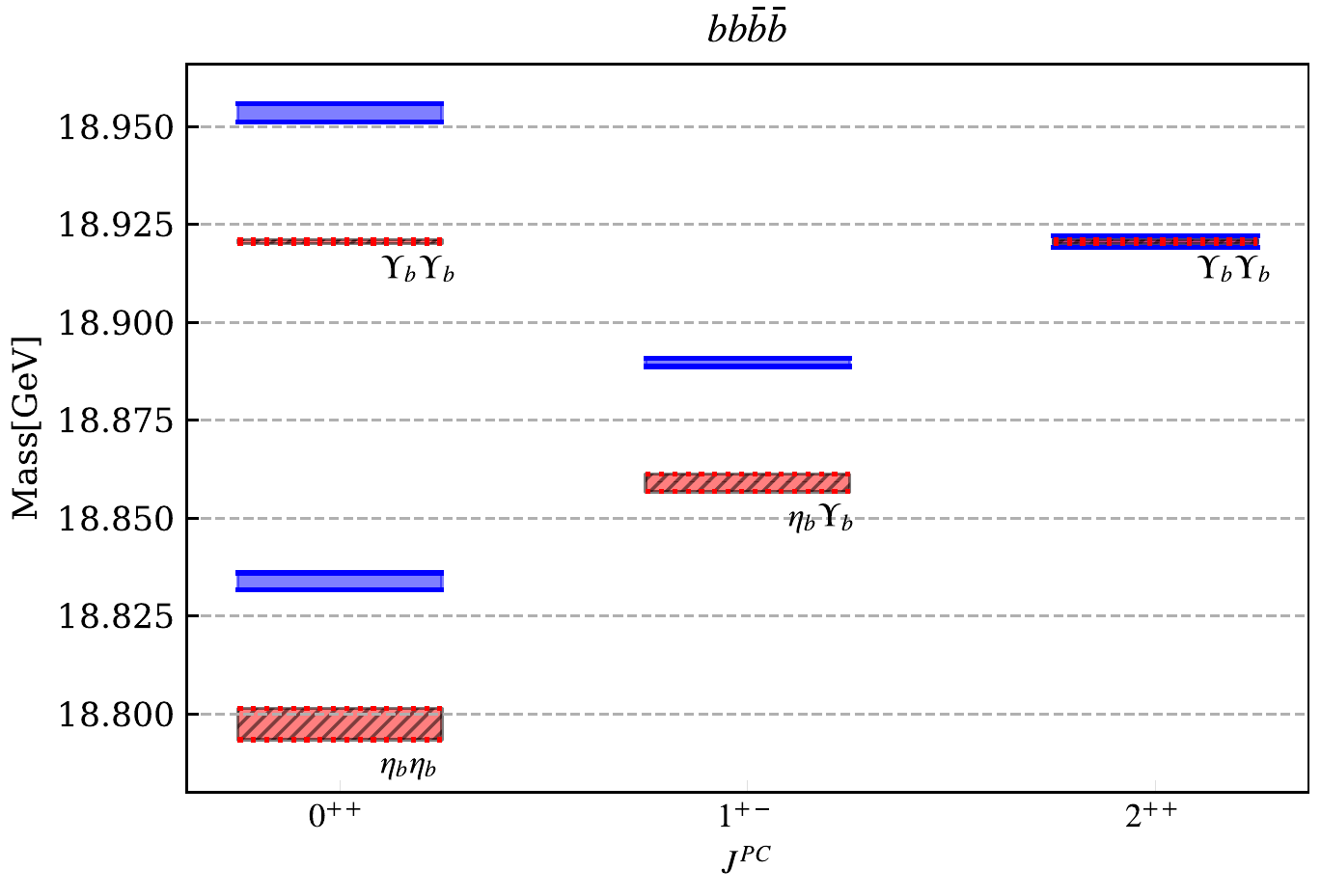}
    \includegraphics[width=70mm]{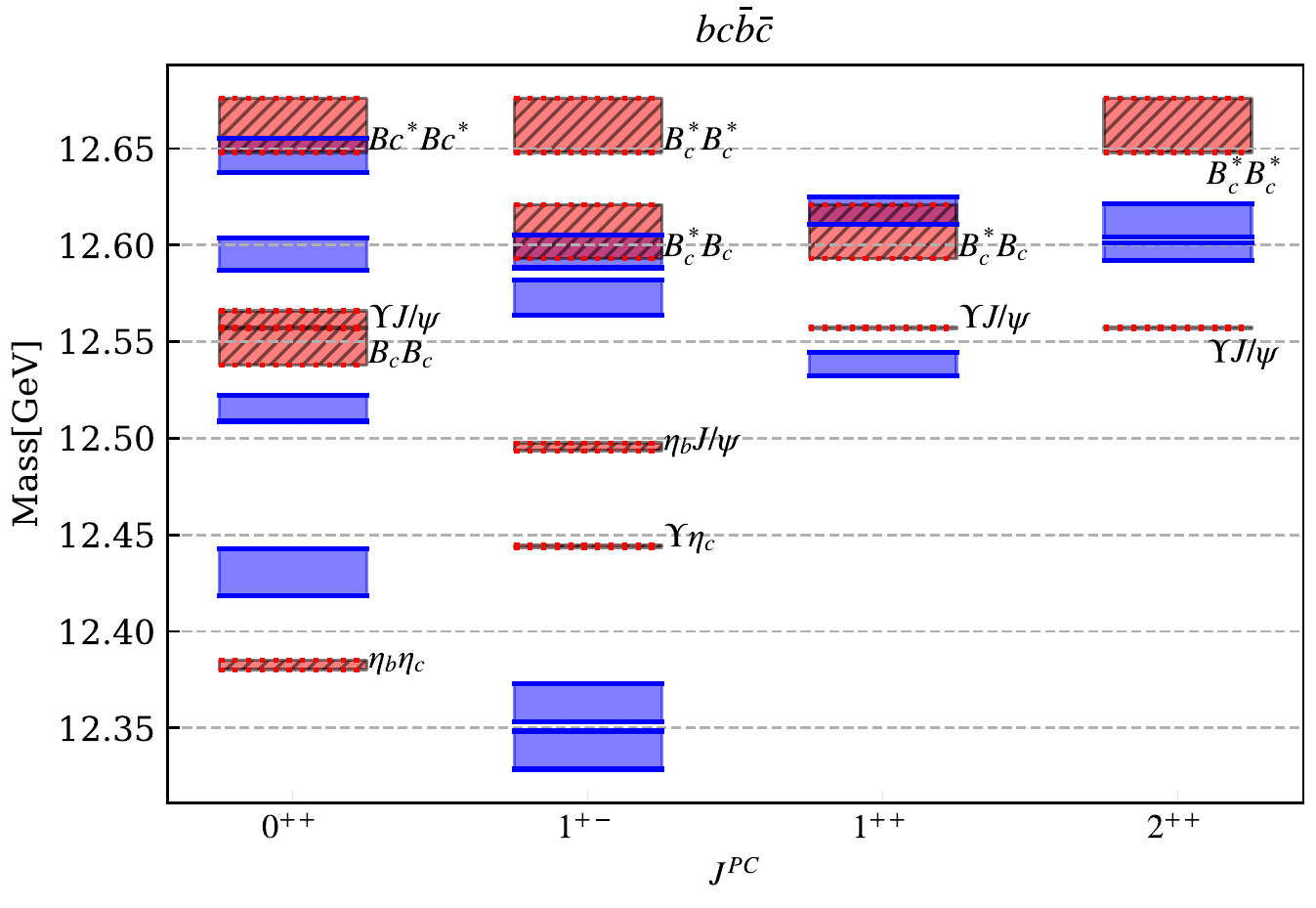}
    \caption{\label{fig:full_heavy_tetra_fig}Mass spectra (blue solid boxes) of $cc\bar{c}\bar{c}$, $bb\bar{b}\bar{b}$ and $bc\bar{b}\bar{c}$ tetraquarks
        for different $J^{PC}$s and their potential hidden charm decay channels (red dashed boxes).  The bands
        are uncertainties either from the framework or the experimental data.
        The green and yellow points are the poles after the coupled channel effect,
        which will be discussed in the next sections. The errors inherit from the experimental data.}
\end{figure}

With all the parameters extracted from the $S$-wave ground pseudoscalar and vector
heavy mesons, one can obtain the matrix elements of Hamiltonian, i.e., Eq.~(\ref{eq:hamil})
in the bases listed in Table~\ref{tab:wave_func_tetra}.  The numerical matrixes of Hamiltonian for various systems
can be found in the appendix.
After diagonalizing Hamiltonian, one can obtain the mass spectra
and eigenvectors which can be found in the appendix. To get an intuitive impression
of the bare tetraquarks, we also plot
the mass spectra of $cc\anticc$, $bb\antibb$, $bc\antibc$ and
their potential hidden charm decay channels in Fig.~\ref{fig:full_heavy_tetra_fig}.
One can see that most of them are above the lowest allowed
decay channels and are expected to illustrate themselves as broader structures.  
However, this expectation might be invaded due to the couplings of the decay channels as discussed below. 

\newcommand{\CoppH}{$\begin{bmatrix}6179.68 & -103.80\\-103.80 &  6109.05\end{bmatrix}$}
\newcommand{\CoppM}{$\begin{bmatrix}6034.72 \\  6254.00\end{bmatrix}$}
\newcommand{\CoppMerr}{$\begin{bmatrix}0.52 \\  0.57\end{bmatrix}$}
\newcommand{\CoppV}{$\begin{bmatrix}0.58 &  0.81\\-0.81 &  0.58\end{bmatrix}$}
\newcommand{\BoppH}{$\begin{bmatrix}18912.91 & -56.58\\-56.58 &  18874.41\end{bmatrix}$}
\newcommand{\BoppM}{$\begin{bmatrix}18833.90 \\  18953.43\end{bmatrix}$}
\newcommand{\BoppMerr}{$\begin{bmatrix}2.12 \\  2.34\end{bmatrix}$}
\newcommand{\BoppV}{$\begin{bmatrix}0.58 &  0.81\\-0.81 &  0.58\end{bmatrix}$}
\newcommand{\BCOppH}{$\begin{bmatrix}12463.09 & 0 & 0 & -65.36\\0 & 12579.53 & -65.36 & 0\\0 & -65.36 & 12582.44 & 0\\-65.36 & 0 & 0 & 12563.01\end{bmatrix}$}
\newcommand{\BCOppM}{$\begin{bmatrix}12515.61 \\ 12646.36 \\ 12430.79 \\ 12595.32\end{bmatrix}$}
\newcommand{\BCOppMerr}{$\begin{bmatrix}6.75 \\ 8.94 \\ 12.14 \\ 8.41\end{bmatrix}$}
\newcommand{\BCOppV}{$\begin{bmatrix} 0 & 0.71 & 0.70 & 0\\ 0 & -0.70 & 0.71 & 0 \\ 0.90 & 0 & 0 & 0.44 \\ -0.44 & 0 & 0 & 0.90 \end{bmatrix}$}
\newcommand{\BCIpmH}{$\begin{bmatrix}12338.56 & 0 & 0 & 0\\0 & 12369.28 & 0 & -37.73\\0 & 0 & 12572.73 & 0\\0 & -37.73 & 0 & 12590.51\end{bmatrix}$}
\newcommand{\BCIpmM}{$\begin{bmatrix}12363.02 \\  12596.77 \\  12338.56 \\  12572.72\end{bmatrix}$}
\newcommand{\BCIpmMerr}{$\begin{bmatrix}9.88 \\  8.49 \\  9.87 \\  9.10\end{bmatrix}$}
\newcommand{\BCIpmV}{$\begin{bmatrix} 0 & 0.99  & 0 & 0.16 \\
                0 & -0.16 & 0 & 0.99 \\
                1 & 0     & 0 & 0    \\
                0 & 0     & 1 & 0\end{bmatrix}$}
\newcommand{\BCIppH}{$\begin{bmatrix}12565.78 & 37.73\\37.73 & 12590.51\end{bmatrix}$}
\newcommand{\BCIppM}{$\begin{bmatrix}12538.44 \\ 12617.85\end{bmatrix}$}
\newcommand{\BCIppMerr}{$\begin{bmatrix}6.14 \\ 7.00\end{bmatrix}$}
\newcommand{\BCIppV}{$\begin{bmatrix} -0.81 & 0.59 \\ 0.59 & 0.81\end{bmatrix}$}
\newcommand{\BCIIppH}{$\begin{bmatrix}12596.50 & 0\\0 & 12612.83\end{bmatrix}$}
\newcommand{\BCIIppM}{$\begin{bmatrix}12596.50 \\ 12612.83\end{bmatrix}$}
\newcommand{\BCIIppMerr}{$\begin{bmatrix}4.57 \\ 8.71\end{bmatrix}$}
\newcommand{\BCIIppV}{$\begin{bmatrix} 1 & 0\\ 0 & 1 \end{bmatrix}$}

\subsection{\label{sec:partial_dw}Partial decay width of bare fully heavy tetraquarks}
As discussed in the above section, most of the bare fully heavy tetraquarks
are above their lowest allowed hidden charm/bottom decay channels~\cite{Ader:1981db,Zhao:2020zjh,Wu:2016vtq,Faustov:2021hjs,Lu:2020cns,Karliner:2016zzc,Deng:2020iqw,Chen:2016jxd,Zhang:2020xtb,Yang:2020wkh,Chen:2020xwe,Zhao:2020nwy,Ke:2021iyh,Zhu:2020xni,Karliner:2020dta,Wang:2019rdo,Weng:2020jao,Berezhnoy:2011xn,Chen:2020lgj}.
As the result, we will discuss their partial widths of hidden charm/bottom
channels. As our framework considers all the possible two-body interactions
among the four constituents,  the $\diq\otimes\antidiq$ base is equal to the
$Q_{1}\bar{Q}_{3}\otimes Q_{2}\bar{Q}_{4}$(or $Q_{1}\bar{Q}_{4}\otimes Q_{2}\bar{Q}_{3}$)
base.  These two bases can be transformed to each other by
Fierz rearrangement~\cite{Becchi:2020mjz,Becchi:2020uvq,Ali:2019roi}
\begin{eqnarray}\label{eq:fierz01}\nonumber
    \ket{\{cc\}_{0}^{6}\{\bar{c}\bar{c}\}_{0}^{\bar{6}}}_{0}^{1}
    &=& \frac{1}{2\sqrt{3}}\ket{\{c\bar{c}\}_{0}^{8}\{c\bar{c}\}_{0}^{8}}_{0}^{1} + \frac{1}{2}\ket{[c\bar{c}]_{1}^{8}[c\bar{c}]_{1}^{8}}_{0}^{1}\\\nonumber
    &+& \sqrt{\frac{1}{6}}\ket{[c\bar{c}]_{0}^{1}[c\bar{c}]_{0}^{1}}_{0}^{1} + \sqrt{\frac{1}{2}}\ket{\{c\bar{c}\}_{1}^{1}\{c\bar{c}\}_{1}^{1}}_{0}^{1},\\\nonumber
    \ket{\{cc\}_{1}^{\bar{3}}\{\bar{c}\bar{c}\}_{1}^{3}}_{0}^{1}
    &=& -\sqrt{\frac{1}{2}}\ket{\{c\bar{c}\}_{0}^{8}\{c\bar{c}\}_{0}^{8}}_{0}^{1} + \sqrt{\frac{1}{6}}\ket{[c\bar{c}]_{1}^{8}[c\bar{c}]_{1}^{8}}_{0}^{1} \\\nonumber
    &+& \frac{1}{2}\ket{[c\bar{c}]_{0}^{1}[c\bar{c}]_{0}^{1}}_{0}^{1} - \frac{1}{2\sqrt{3}}\ket{\{c\bar{c}\}_{1}^{1}\{c\bar{c}\}_{1}^{1}}_{0}^{1},\\\nonumber
    \ket{\{cc\}_{1}^{\bar{3}}\{\bar{c}\bar{c}\}_{1}^{3}}_{1}^{1}
    &=&-\sqrt{\frac{1}{3}}\ket{\{c\bar{c}\}_{0}^{8}[c\bar{c}]_{1}^{8}}_{1}^{1} - \sqrt{\frac{1}{3}}\ket{[c\bar{c}]_{1}^{8}\{c\bar{c}\}_{0}^{8}}_{1}^{1} \\\nonumber
    &+& \sqrt{\frac{1}{6}}\ket{[c\bar{c}]_{1}^{1}\{c\bar{c}\}_{0}^{1}}_{1}^{1} + \sqrt{\frac{1}{6}}\ket{\{c\bar{c}\}_{0}^{1}[c\bar{c}]_{1}^{1}}_{1}^{1},\\\nonumber
    \ket{\{cc\}_{1}^{\bar{3}}\{\bar{c}\bar{c}\}_{1}^{3}}_{2}^{1}
    &=&\sqrt{\frac{1}{3}}\ket{\{c\bar{c}\}_{1}^{1}\{c\bar{c}\}_{1}^{1}}_{2}^{1}-\sqrt{\frac{2}{3}}\ket{[c\bar{c}]_{1}^{8}[c\bar{c}]_{1}^{8}}_{2}^{1}.
\end{eqnarray}
Here we use the $cc\bar{c}\bar{c}$ system as an example and the
transformation for other systems are analogous. Because the observed
hadrons are color singlet, only the $\ket{(c\bar{c})^{1}(c\bar{c})^{1}}^{1}$ components
on the right hand side of the above equations contribute to the hidden charm decay channels.
Here $(c\bar{c})$ means either $\{c\bar{c}\}$ or $[c\bar{c}]$ which depends on
the spin of the $c\bar{c}$ pair. The coefficient of the $\ket{(c\bar{c})_{s_{i1}}^{1}(c\bar{c})_{s_{i2}}^{1}}^{1}_{J^{PC}_n}$ component
is denoted as $\alpha_n^i$.  Here $i$ indicates the allowed two-body charmonium decay channels.
$s_{i1}$ and $s_{i2}$ are the spins of the two $c\bar{c}$ pairs in the $i$-th channel.
The subscript $J^{PC}_n$ means the $n$-th $J^{PC}$ base.
The two physical $0^{++}$ tetraquarks
are a combination of the first two bases with the mixing coefficients
listed in the last column Table~\ref{tab:full_heavy_tetra_tab} of the appendix.
Collect these coefficients into a matrix as
\begin{eqnarray}
    \beta=\left(\begin{array}{cc}
        0.58  & 0.81 \\
        -0.81 & 0.58
    \end{array}\right).
\end{eqnarray}

With all the pieces ready, the transition rates of physical tetraquarks to two hidden charmonium channels
can be read through the coefficient matrixes
\begin{eqnarray}\nonumber
    \mu( X_{0^{++}}(6035))&=&\left(\begin{array}{cc}
            \sqrt{\frac{1}{6}}\beta_{11}+\frac{1}{2}\beta_{12} & \sqrt{\frac{1}{2}}\beta_{11}-\frac{1}{2\sqrt{3}}\beta_{12}
        \end{array}\right)   \\\nonumber
    \mu( X_{0^{++}}(6254))&=&\left(\begin{array}{cc}
            \sqrt{\frac{1}{6}}\beta_{21}+\frac{1}{2}\beta_{22} & \sqrt{\frac{1}{2}}\beta_{21}-\frac{1}{2\sqrt{3}}\beta_{22}
        \end{array}\right)
\end{eqnarray}
in the $\ket{[c\bar{c}]_{0}^{1}[c\bar{c}]_{0}^{1}}_{0}^{1}$, $\ket{\{c\bar{c}\}_{1}^{1}\{c\bar{c}\}_{1}^{1}}_{0}^{1}$ base,
\begin{eqnarray}\nonumber
    \mu(X_{1^{+-}}(6137))=\left(\begin{array}{cc}
            \sqrt{\frac{1}{6}} & \sqrt{\frac{1}{6}}\end{array}\right)
\end{eqnarray}
in the $\ket{[c\bar{c}]_{1}^{1}\{c\bar{c}\}_{0}^{1}}_{1}^{1}$, $\ket{\{c\bar{c}\}_{0}^{1}[c\bar{c}]_{1}^{1}}_{1}^{1}$ base, and
\begin{eqnarray}\nonumber
    \mu(X_{2^{++}}(6194))=\ensuremath{\sqrt{\frac{1}{3}}}
\end{eqnarray}
in the $\ket{\{c\bar{c}\}_{1}^{1}\{c\bar{c}\}_{1}^{1}}_{2}^{1}$ base.
By expanding explicitly, one obtains
\begin{widetext}
    \begin{eqnarray}\label{eq:wf1}
        X_{0^{++}}(6035)	&\sim&\left(\sqrt{\frac{1}{6}}\beta_{11}+\frac{1}{2}\beta_{12}\right)\ket{[c\bar{c}]_{0}^{1}[c\bar{c}]_{0}^{1}}_{0}^{1}+\left(\sqrt{\frac{1}{2}}\beta_{11}-\frac{1}{2\sqrt{3}}\beta_{12}\right)\ket{\{c\bar{c}\}_{1}^{1}\{c\bar{c}\}_{1}^{1}}_{0}^{1},\\\label{eq:wf2}
        X_{0^{++}}(6254)	&\sim&\left(\sqrt{\frac{1}{6}}\beta_{21}+\frac{1}{2}\beta_{22}\right)\ket{[c\bar{c}]_{0}^{1}[c\bar{c}]_{0}^{1}}_{0}^{1}+\left(\sqrt{\frac{1}{2}}\beta_{21}-\frac{1}{2\sqrt{3}}\beta_{22}\right)\ket{\{c\bar{c}\}_{1}^{1}\{c\bar{c}\}_{1}^{1}}_{0}^{1},\\\label{eq:wf3}
        X_{1^{+-}}(6137)	&\sim& \sqrt{\frac{1}{6}}\ket{[c\bar{c}]_{1}^{1}\{c\bar{c}\}_{0}^{1}}_{1}^{1} + \sqrt{\frac{1}{6}}\ket{\{c\bar{c}\}_{0}^{1}[c\bar{c}]_{1}^{1}}_{1}^{1},\\\label{eq:wf4}
        X_{2^{++}}(6194)	&\sim& \sqrt{\frac{1}{3}}\ket{\{c\bar{c}\}_{1}^{1}\{c\bar{c}\}_{1}^{1}}_{2}^{1},
    \end{eqnarray}
\end{widetext}
with their masses in $(\cdots)$. Here the values in $(\cdots)$ are
the bare pole masses before the coupled channel effect which
will be discussed in the next section.
With the definitons
\begin{eqnarray}\label{eq:had1}
    g_{\text{di}-J/\psi}&\defeq&\langle J/\psi J/\psi|\hat{H}_\text{strong}|\ket{\{c\bar{c}\}_{1}^{1}\{c\bar{c}\}_{1}^{1}}_{J}^{1}\rangle,\\\label{eq:had2}
    g_{\text{di}-\eta_c}&\defeq&\langle \eta_c\eta_c|\hat{H}_\text{strong}|\ket{[c\bar{c}]_{0}^{1}[c\bar{c}]_{0}^{1}}_{J}^{1}\rangle
\end{eqnarray}
indicating the hadronization process,
one can obtain the relative transition rate to
di-$J/\psi$ is
\begin{widetext}
    \begin{eqnarray}\nonumber
        &&  |\mathcal{M}(X^{\text{di}-J/\psi}_{2^{++}}(6194) )|^2:|\mathcal{M}(X^{\text{di}-J/\psi}_{0^{++}}(6254) )|^2: |\mathcal{M}(X^{\text{di}-J/\psi}_{0^{++}}(6035) )|^2\\
        &=&\frac{1}{3}g_{\text{di}-J/\psi}^2:\left(\sqrt{\frac{1}{2}}\beta_{21}-\frac{1}{2\sqrt{3}}\beta_{22}\right)^2g_{\text{di}-J/\psi}^2:\left(\sqrt{\frac{1}{2}}\beta_{11}-\frac{1}{2\sqrt{3}}\beta_{12}\right)^2g_{\text{di}-J/\psi}^2\sim 33:55:3
    \end{eqnarray}
\end{widetext}
and that to di-$\eta_c$ is
\begin{widetext}
    \begin{eqnarray}\nonumber
        && |\mathcal{M}(X^{\text{di}-\eta_c}_{0^{++}}(6254) )|^2: |\mathcal{M}(X^{\text{di}-\eta_c}_{0^{++}}(6035) )|^2\\&=&\left(\sqrt{\frac{1}{6}}\beta_{21}+\frac{1}{2}\beta_{22}\right)^2g_{\text{di}-\eta_c}^2:\left(\sqrt{\frac{1}{6}}\beta_{11}+\frac{1}{2}\beta_{12}\right)^2g_{\text{di}-\eta_c}^2\sim 0:41.
    \end{eqnarray}
\end{widetext}
Although the hadronization parameters does not play a role here,
it will when the coupled channel effect is included.
Considering the $S$-wave phase space
\begin{equation}\label{eq:two-body decay}
    \text{P.S.} = \frac{1}{8\pi}  \frac{|\bm{p}|}{M^{2}}
\end{equation}
with $|\bm{p}|$ the three momentum of the final particle in the rest frame of the decaying tetraquark,
the ratio to di-$\eta_c$ becomes $1:127$. It means that the lower $X_{0^{++}}(6035)$
should be more significant in di-$\eta_c$ channel,
which is also the case after the coupled channel effect.
After inclusion of the $S$-wave phase space, only the
$X_{0^{++}}(6254)$ is allowed to decay into di-$J/\psi$
and would be expected to be significant in the di-$J/\psi$ line shape.
The values for other systems can be found in the appendix.

\subsection{\label{sec:coupled_channel} Coupled channel effect to two hidden charmonium channels}
From Eqs.~\eqref{eq:wf1},\eqref{eq:wf2},\eqref{eq:wf3},\eqref{eq:wf4} and Eqs.~\eqref{eq:had1},\eqref{eq:had2}
one also can obtain the potentials between two hidden charmonium channels via bare tetraquark poles in the
above subsection. As the allowed C-parity of di-$J/\psi$ system
is positive, only the $0^{++}$ and $2^{++}$ quantum numbers could explain the
structure in experiment~\cite{LHCb:2020bwg}. For the $0^{++}$ channel, we consider
the $\eta_c\eta_c$, $J/\psi J/\psi$, $J/\psi\psi^\prime$, $J/\psi\psi(3770)$
channels. The inclusion of the latter two is because the significant changes in the line shape.
The corresponding $0^{++}$ potential reads as
\begin{eqnarray}\label{eq:potential}
    V^{0^{++}}_{ij}(E) = \sum_{n=1,2}\sum_{\alpha,\beta=1,2} \frac{\mu_{n}^{\alpha}\mu_{n}^{\beta}g^\alpha_{i}g^\beta_{j}}{E-E_{n0}}
\end{eqnarray}
with $E_{10}$ and $E_{20}$ the masses of bare compact fully heavy tetraquarks
$ X_{0^{++}}(6035)$ and $ X_{0^{++}}(6254)$, respectively.
$\mu_n^\alpha$ is the coefficient of the $\ket{[c\bar{c}]_{0}^{1}[c\bar{c}]_{0}^{1}}_{0}^{1}$,
$\ket{\{c\bar{c}\}_{1}^{1}\{c\bar{c}\}_{1}^{1}}_{0}^{1}$ components of the n-th compact tetraquarks.
$g^\alpha_i$ represents the $\alpha$-th component hadronization to the $i$-th channel (analogous to Eqs.~\eqref{eq:had1},\eqref{eq:had2}).
For the $2^{++}$ channel, we have $J/\psi J/\psi$, $J/\psi\psi^\prime$, $J/\psi\psi(3770)$
channels and the corresponding potential can be obtained analogous to Eq.\eqref{eq:potential}.
\begin{table}[!htbp]
    \caption{\label{tab: parameters} The values of the parameters extracted from the fit.
    The errors are from the experimental uncertainties. The subscripts are for the
    corresponding channels. }
    \begin{tabular}{|c|c|c|}
\hline 
parameters & $0^{++}$ & $2^{++}$\tabularnewline
\hline 
\hline 
$U_{\eta_{c}\eta_{c}}^{J^{PC}}$ & $-572.92\pm912.33$ & --\tabularnewline
\hline 
$U_{J/\psi J/\psi}^{J^{PC}}$ & $7.53\pm3.87$ & $30.67\pm1.93$\tabularnewline
\hline 
$U_{J/\psi\psi^{\prime}}^{J^{PC}}$ & $34447.71\pm4145.37$ & $39111.96\pm6605.72$\tabularnewline
\hline 
$U_{J/\psi\psi^{\prime\prime}}^{J^{PC}}$ & $-37513.64\pm4035.80$ & $-51446.38\pm7192.27$\tabularnewline
\hline 
$g_{J/\psi J/\psi}$ & \multicolumn{2}{c|}{$0.989\pm0.03$}\tabularnewline
\hline 
$g_{\eta_{c}\eta_{c}}$ & \multicolumn{2}{c|}{$0.924\pm0.03$}\tabularnewline
\hline 
$g_{J/\psi\text{\ensuremath{\psi^{\prime}}}}$ & \multicolumn{2}{c|}{$0.177\pm0.02$}\tabularnewline
\hline 
$g_{J/\psi\psi^{\prime\prime}}$ & \multicolumn{2}{c|}{$0.134\pm0.01$}\tabularnewline
\hline 
\end{tabular}
\end{table}
    \begin{table*}[!htbp]
        \caption{\label{tab:poles} The pole positions of the $0^{++}$ and $2^{++}$ channels
            comparing with the bare tetraquark masses in the bracket.
            The errors of pole positions inherit from the experimental errors of the di-$J/\psi$ invariant mass distribution.}
        \begin{tabular}{@{}cccc@{}}
            \hline\hline
            {}                     & $0^{++}$                                                                       & {} & $2^{++}$                                              \\\hline
            \multirow{2}{*}{Poles (MeV)}  & $6251.65^{+5}_{-54}+1.47^{+0.06}_{-1.12}i$  ($6254^{+0.57}_{-0.57}$)           & {} & $6192.46^{+1.23}_{-12.56}$ ($6193.8^{+0.35}_{-0.35}$) \\
                                   & $6032.96^{+60.16}_{-21.82}+1.036^{+0.60}_{-0.31}i$ ($6034.72^{+0.52}_{-0.52}$) & {}                                                         \\
            \hline\hline
        \end{tabular}
    \end{table*}

\begin{figure}[htbp]
    \centering
    \includegraphics[width=70mm]{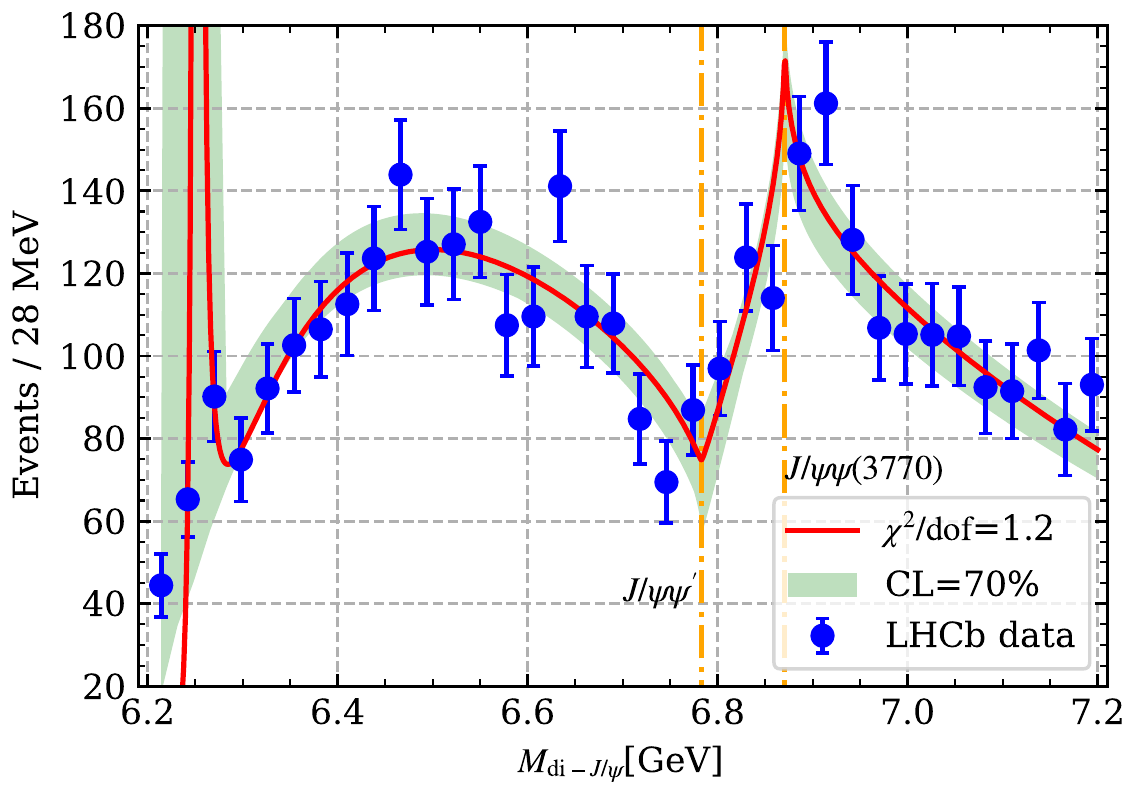}
    \caption{\label{fig:fit}The fitted result (red solid curve) comparing
        to the experimental data~\cite{LHCb:2020bwg}.
        The band is the error with $70\%$ confidence level.
        The two orange vertical lines are the $J/\psi\psi^\prime$ and the $J/\psi\psi(3770)$ thresholds. }
\end{figure}

\begin{figure}
    \centering
    \includegraphics[width=70mm]{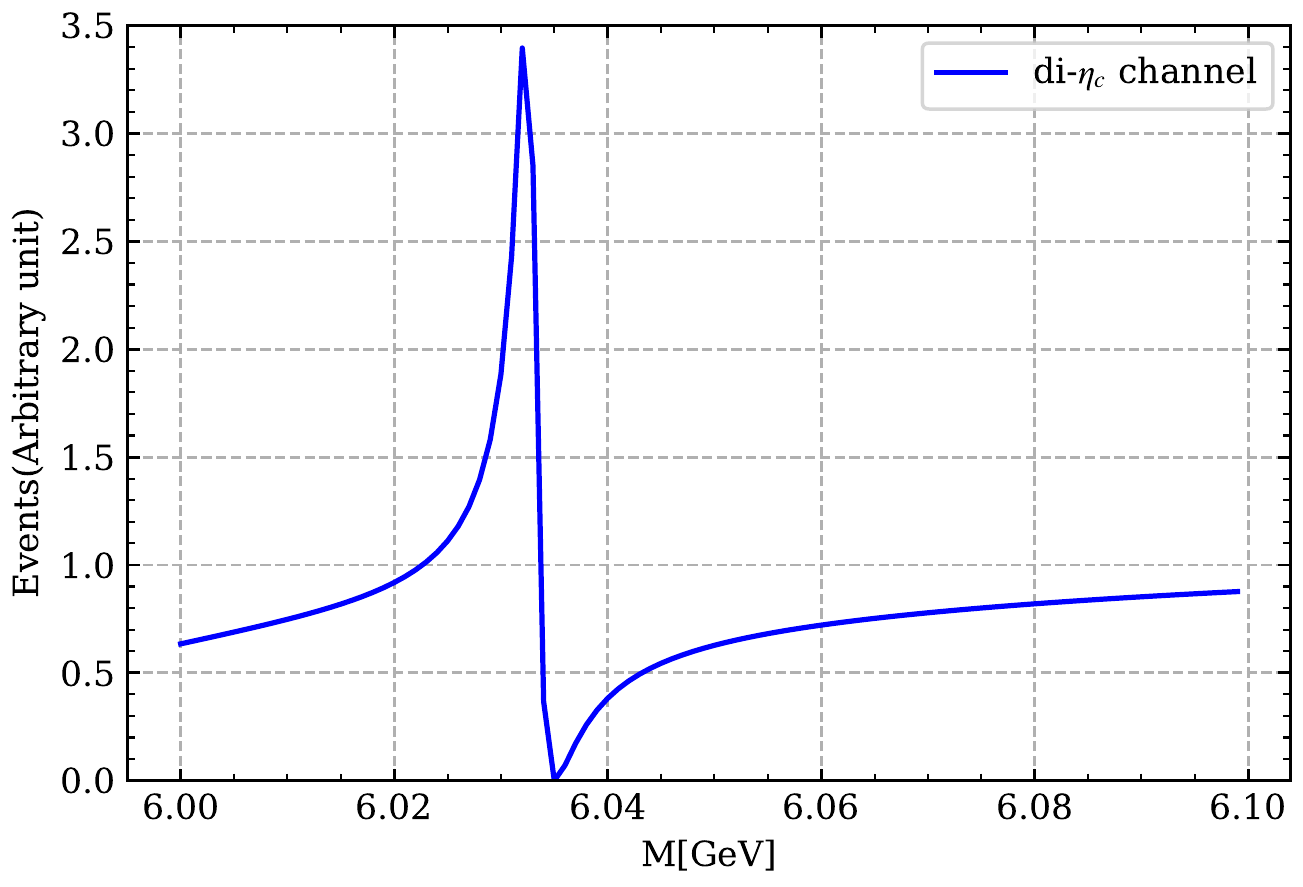}
    \caption{\label{fig:dietac} 
        Prediction of the di-$\eta_{c}$ lineshape.}
\end{figure}

The two-body propagator for the $i$-th channel is
\begin{equation}
    \begin{aligned}\label{eq:G}
        G_{i}(E) & = \frac{1}{16\pi^{2}}\Big\{a(\mu)+\log\frac{m_{i1}^{2}}{\mu^2}+\frac{m_{i2}^2-m_{i1}^2+s}{2s}\log
        \frac{m_{i2}^{2}}{m_{i1}^{2}}                                                                                \\
                 & +\frac{k}{E}\Big[\log\bk{2k_{i}E+s+\Delta_{i}}+\log\bk{2k_{i}E+s-\Delta_{i}}                      \\
                 & -\log\bk{2k_{i}E-s+\Delta_{i}}-\log(2k_{i}E-s-\Delta_{i})\Big]\Big\}
    \end{aligned}
\end{equation}
where $s=E^{2}$, $m_{i1}$ and $m_{i2}$ are the particle masses in the \emph{i}th channel, $\Delta_{i}=m_{i1}^{2}-m_{i2}^{2}$, $k_{i}=\lambda^{1/2}(E^{2},m_{i1}^{2},m_{i2}^{2})/2E$.
Here the dimensional regularization is used to regularize the ultra-divergence and we take $a(\mu)=-3$, $\mu=1\gev$. The physical scattering T matrix can be obtained by solving Lippmann Schwinger Equation
\begin{eqnarray}
    T=V+VGT.
\end{eqnarray}
The physical production amplitude of double $J/\psi$ is
\begin{widetext}
    \begin{equation}\label{eq:P0pp}
        P_{2}^{0^{++}} = U^{0^{++}}_{2} + U^{0^{++}}_{1}G_{1}T_{12} + U^{0^{++}}_{2}G_{2}T_{22}
        + U^{0^{++}}_{3}G_{3}T_{32} + U^{0^{++}}_{4}G_{4}T_{42}
    \end{equation}
\end{widetext}
where $U^{0^{++}}_i$ stands for the bare production amplitude for the $i$-th channel with $J^{PC}=0^{++}$.
Notice that the channels $\eta_{c}\eta_{c}$, $J/\psi J/\psi$, $J\psi\psi'$ and $J\psi\psi(3770)$ are ordered by their thresholds.
Analogously, the physical production amplitude of di-$J/\psi$ for the $2^{++}$ state is
\begin{equation}\nonumber
    P_{1}^{2^{++}} = U^{2^{++}}_{1} + U^{2^{++}}_{1}G_{1}T_{11} + U^{2^{++}}_{2}G_{2}T_{21} + U^{2^{++}}_{3}G_{3}T_{31}.
\end{equation}
Here, $G_{i}$ representes the two-loop function of $J/\psi J/\psi$,  $J\psi\psi'$ and $J\psi\psi(3770)$
and $U^{2^{++}}_i$ stands for the bare production amplitude for the $i$-th channel with $J^{PC}=2^{++}$.
We adopt the $S$-wave phase space factor
\begin{equation}
    \rho(E) = \frac{|{\bm k}|}{8\pi E}.
\end{equation}
with $|{\bm k}|$ the three momentum of the particle in the center-of-rest frame.
similar to Refs~\cite{Dong:2020nwy}. The final fit function is
\begin{equation}
    \bk{\left|P_{2}^{{0^{++}}}\right|^{2}+\left|P_{1}^{2^{++}}\right|^{2}}\rho(E).
\end{equation}
The fitted di-$J/\psi$ invariant mass distribution comparing to the experimental data
is shown in Fig.~\ref{fig:fit}, with the fitted parameters in Table~\ref{tab: parameters}.
From the figure, one can see a significant structure around $6.25~\mathrm{GeV}$
which might stem from the shifted $X_{0^{++}}(6254)$. This structure
could be seen when the experimental statistic increases
and can be viewed as a strong evidence of the compact fully heavy tetraquarks.
 The structure around $6.9~\mathrm{GeV}$
demonstrates itself as a cusp effect from the $J/\psi\psi(3770)$ channel.
As stated in above section, the bare $X_{0^{++}}(6035)$ strongly couples to the di-$\eta_c$ channel
and will demonstrate itself in this channel. In addition,
even after the coupled channel effect, the shift of this bare
state is marginal. That means the results for the bare pole
almost survive and the 
$X_{0^{++}}(6035)$ will also show a significant peak structure in 
the di-$\eta_c$ spectrum, which 
is a key physical observable for the nature of the fully heavy tetraquark.
For the further measurement in experiment,
the di-$\eta_c$ line shape
is presented in Fig.~\ref{fig:dietac} without any background.

\begin{figure}[!htbp]
    \centering
    \includegraphics[width=80mm]{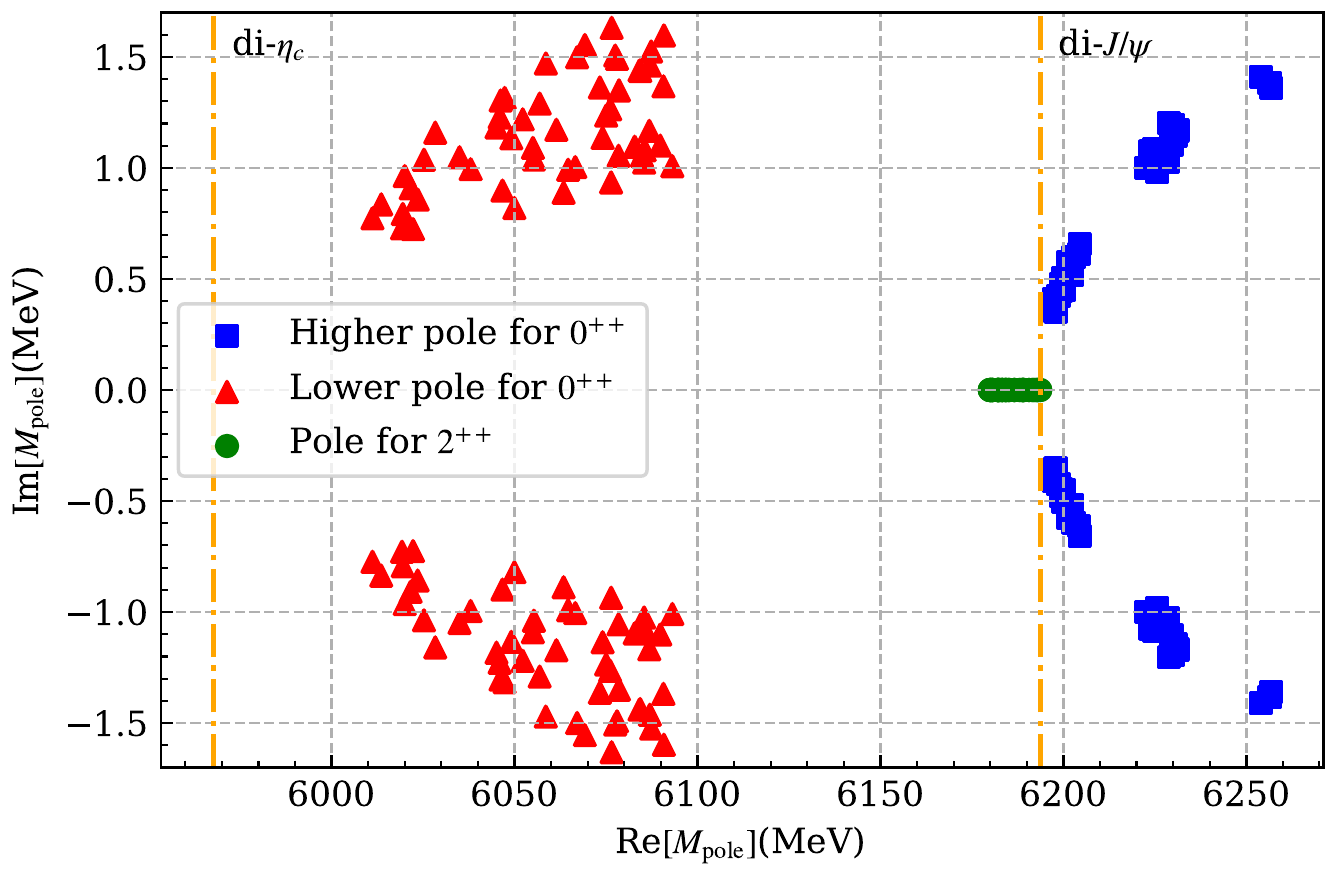}
    \caption{\label{fig:pole positions}Pole positions of the $0^{++}$
        (red triangles and blue boxes for the lower and higher ones, respectively) and $2^{++}$ (green circles) channels
        with parameters in $70\%$ confidence level.
        The yellow dashed vertical lines are the di-$\eta_c$ and di-$J/\psi$
        channels, respectively. }
\end{figure}

To further check this understanding, the pole positions of the corresponding $S$-matrix
are extracted and shown in Table~\ref{tab:poles} comparing to the bare pole masses~
\footnote{The pole positions have also been extracted in Refs.~\cite{Dong:2020nwy,Liang:2021fzr,Guo:2020pvt} based on the molecular picture, which might have different quantum numbers comparing with the compact ones here. As the result, the comparison with their results are not presented.}.
One can see that the shift of the pole positions to the bare masses are
tiny which indicates that the physical states are dominated by compact tetraquarks.
We do not find a pole around $6.9~\mathrm{GeV}$ which indicates the structure is only the 
cusp effect from the $J/\psi\psi(3770)$ channel. One can also see a dip structure in Fig.~\ref{fig:fit}
around the $J/\psi\psi^\prime$ threshold. 
The pole positions with the parameters within $70\%$ confidence level
are shown in Fig.~\ref{fig:pole positions}. The two $0^{++}$ tetraquarks
behave as two resonances above the di-$\eta_c$ and di-$J/\psi$ threshold, respectively.
The $2^{++}$ state is a bound state below the di-$J/\psi$ threshold. 
The positions of these three states are listed in Table~\ref{tab:poles}
comparing with the bare pole masses. 
In addition, the nature of full-charm tetraquark states
can also be obtained by estimating their compositeness
$\bar{X}_{A}=1-\bar{Z_A}$~\cite{Dong:2020nwy,Matuschek:2020gqe} with $\bar{Z}_{A}=1$ for
molecules and $\bar{Z}_{A}=0$ for compact states, respectively.
This idea was proposed by Weinberge in 1963 for bound states~\cite{Weinberg:1962hj}
and extended to virtual states and resonances~\cite{Baru:2003qq,Matuschek:2020gqe} recently.
Practically, the di-$J/\psi$ to di-$J/\psi$ scattering amplitude
can be parametrized in the effective range expansion
\begin{equation}\label{eq:SA}
    T(k) = -8\pi\sqrt{s}\left[\frac{1}{a_{0}}+\frac{1}{2}r_{0}k^{2}-ik+\mathcal{O}(k^{4})\right]^{-1},
\end{equation}
with $a_0$, $r_0$ the scattering length and effective range, respectively. 
$k$ is the three momentum of the $J/\psi$ in the di-$J/\psi$ center-of-mass frame.
By comparing with the physical scattering amplitude,
one can extract the S-wave scattering length $a_{0}$ and the effective range $r_{0}$.
Furthermore, one can obtain the compositeness~\cite{Matuschek:2020gqe}
\begin{equation}\label{eq:XA}
    \bar{X}_{A} = \frac{1}{\sqrt{1+2|r_{0}/a_{0}|}}.
\end{equation}
The results are collected in Table \ref{tab:compositieness}.
The absolute values of effective range are much larger than
those of the scattering lengths for both $0^{++}$ and $2^{++}$
channels. That indicates that compact tetraquarks are the
dominant contributions which can also be understood by the large values of 
$\bar{Z}_A$ in Table \ref{tab:compositieness}. This behavior has also been 
explained by the frame with the so-called Castillejo-Dalitz-Dyson (CDD) pole
~\cite{Guo:2016wpy,Kang:2016ezb,Guo:2020pvt},
which state that the scattering length $a_0$ and effective range $r_0$
should be linearly and quadratically inversely proportional to the  distance
 between the CDD pole $M_\text{CDD}$ and the relevant threshold $m_\text{thr.}$, respectively, i.e.
 \begin{eqnarray}
 a_0\propto M_\text{CDD}-m_\text{thr.},\quad r_0\propto \left(M_\text{CDD}-m_\text{thr.} \right)^{-2}.
 \end{eqnarray}
\begin{table}[!htbp]
    \caption{\label{tab:compositieness}
    The scattering length $a_0$, effective range $r_0$ in the di-$J/\psi$ channel 
    as well as the corresponding compositeness $\bar{X}_A$
    and wave function renormalization constants $\bar{Z}_A$
    for the $0^{++}$ and $2^{++}$ channels. The errors are from the 
    uncertainties of the experimental data. }
    \begin{tabular}{@{}llr@{}}
        \hline\hline
        {}            & $0^{++}$                    & $2^{++}$                    \\\hline
        $a_{0}$(fm)   & $0.012_{-5.142}^{+3.129}$    & $-0.280_{-2.397}^{+0.443}$  \\
        $r_{0}$(fm)   & $-37.966_{-4.882}^{+4.010}$ & $-60.803_{-15.222}^{+1.592}$ \\
        $\bar{X}_{A}$ & $0.013_{-0.003}^{+0.241}$   & $0.048_{-0.042}^{+0.095}$   \\
        $\bar{Z}_{A}$ & $0.987_{-0.241}^{+0.003}$   & $0.952_{-0.095}^{+0.042}$
        \\
        \hline\hline
    \end{tabular}
\end{table}
 \begin{figure*}[!htbp]
    \centering
    \includegraphics[scale=.5]{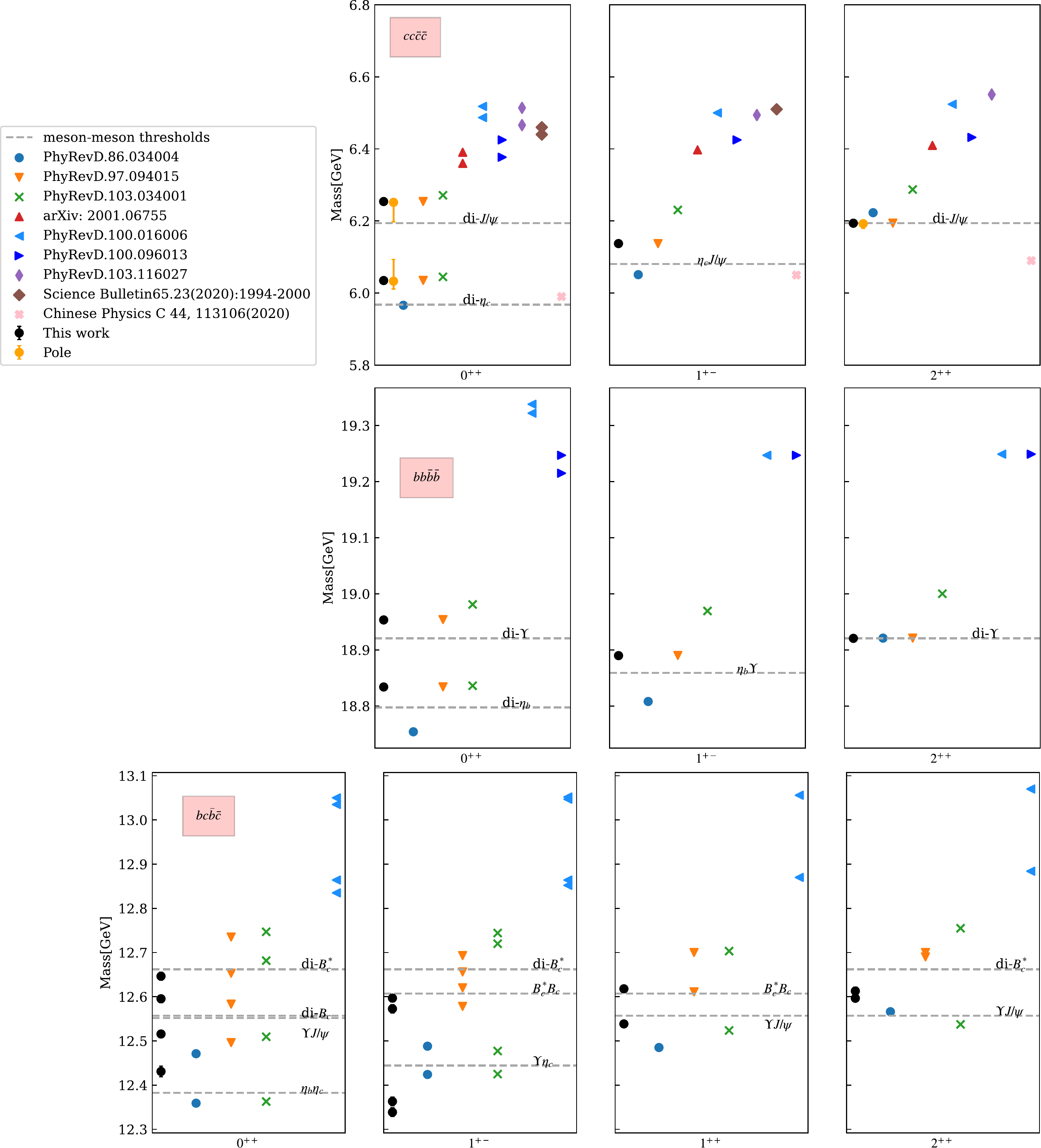}
    \caption{\label{fig:comparsion_fig} Comparison of the fully heavy tetraquark spectra (in units of GeV)
    with other works~\cite{Berezhnoy:2011xn, Wang:2019rdo,Wu:2016vtq, Weng:2020jao, Liu:2019zuc, Zhao:2020zjh, Chen:2020lgj, Wang:2020ols, Chen:2020xwe,Li:2021ygk,Faustov:2020qfm,Bedolla:2019zwg,Tiwari:2021tmz,Yang:2021hrb}. The black circles are for the compact fully heavy tetraquarks. 
    The yellow circles in the $cc\bar{c}\bar{c}$ spectra are the pole positions. 
    The gray dashed lines are for the thresholds of potential two-charmonium decay channels. }
\end{figure*}

To get a whole picture of the fully heavy tetraquark system,
we compare our results with those presented in Refs~\cite{Berezhnoy:2011xn,Wu:2016vtq,Weng:2020jao,Chen:2020lgj,Liu:2019zuc,Wang:2019rdo,Zhao:2020zjh,Chen:2020xwe,Wang:2020ols}
in the compact tetraquark picture.
Fig.~\ref{fig:comparsion_fig} are the comparison of the spectra from these works.
For the full-charm tetraquark system $cc\bar{c}\bar{c}$,
the spectra has been obtained with parameterization scheme in Refs.~\cite{Berezhnoy:2011xn,Wu:2016vtq,Weng:2020jao} and our work, Gaussion expansion method (GEM)~\cite{Hiyama:2012sma} in Refs~\cite{Chen:2020lgj,Liu:2019zuc,Wang:2019rdo,Zhao:2020zjh} and QCD sum rules in Refs.~\cite{Chen:2020xwe,Wang:2020ols}. 
The masses of fully charm tetraquarks mostly fall in the internal $[6.2,6.8]$GeV,
 which are in a good agreement with the range of mass for broad structure investigated by LHCb~\cite{LHCb:2020bwg}. 
 In Ref.~\cite{Berezhnoy:2011xn}, the mass splitting is dominated only by spin interaction.
 Refs.~\cite{Wu:2016vtq,Deng:2020iqw,Liu:2019zoy} further include 
 the chromoelectro and chromomagnetic interaction and find that these two kinds of interaction
 cannot be neglected to obtain the correct spectrum. Among them, our method is similar to that in 
 Ref.~\cite{Weng:2020jao} and our results should be consistent with each other. 
 However, one can see a deviation from Fig.~\ref{fig:comparsion_fig}, which could stem from the
 different methods of extracting the parameters. Our framework is under non-relativistic
 approximation and the kinematic terms have been ignored comparing with the heavy quark mass terms
 and chromoelectro/chromomagnetic interaction. For Refs.~\cite{Weng:2020jao,Weng:2018mmf} 
 both the light meson and heavy meson masses are used to extract the parameters, 
 which is the reason for the deviation. The spectra can also be obtained by solving Schr\"odinger equation
 numerically with the variational method~\cite{Chen:2020lgj,Liu:2019zuc,Wang:2019rdo,Zhao:2020zjh}, which can only give the upper limits of the system.
 It is the reason why the masses with the variational method 
 is much larger than those with the parameterization ones~\cite{Weng:2020jao,Weng:2018mmf} and ours. 
 QCD sum rule can also be used to obtain the spectra~\cite{Chen:2020xwe,Wang:2020ols} of fully heavy system.
 Ref~\cite{Chen:2020xwe} concludes that 
 the broad structure around 6.2-6.8GeV and $X(6900)$ is an $S$-wave
  and a $P$-wave full-charm tetraquark states, respectively.
  On the contrary, Ref.~\cite{Wang:2020ols} shows that the broad structure
   is the first radial excited state of $cc\bar{c}\bar{c}$ tetraquark
   and the $X(6900)$ is the second radial excited state of $cc\bar{c}\bar{c}$ tetraquark. 
   
  \section{Summary}
   In this work, we first extract the internal structure of the fully heavy tetraquarks directly
   from the experimental data, within the compact tetraquark picture. 
   The bare pole masses are obtained from the parametrization with
   both chromoelectro and chromomagnetic interactions. For the $S$-wave ground states,
   the spacial wave function is trivial and their overlapping can be neglected. 
   The rest parameters are extracted from the masses of the $S$-wave ground heavy mesons,
   i.e. the $J/\psi$, $\eta_c$, $\Upsilon(1S)$, $\eta_b(1S)$, $B_c$ and $B_c^*$. 
   Most of the bare masses are above the lowest allowed two-heavy-quarkonium decay channels.
   However, it does not mean that all of them could exhibit themselves
   as broader structures in the lineshape. For an illustration, although the 
   $X_{0^{++}}(6035)$ mass is smaller than the $X_{0^{++}}(6254)$, its transition to the di-$\eta_c$
   channel is much larger than that of the $X_{0^{++}}(6254)$. This
   makes the $X_{0^{++}}(6035)$ more significant in the di-$\eta_c$ lineshape even 
   after the coupled channel effect. This is an unique feature
   which can distinguish compact $cc\bar{c}\bar{c}$ tetraquark from the loosely hadronic molecule. 
   After fit to the di-$J/\psi$ lineshape, we find that
the $X(6900)$ reported by LHCb is only a cusp effect from the $J/\psi\psi(3770)$ channel.
In addition, there is also a cusp effect slightly below $6.8~\mathrm{GeV}$ stemming from the $J/\psi\psi^\prime$ channel. 
The two $0^{++}$ tetraquarks
behave as two resonances above the di-$\eta_c$ and di-$J/\psi$ threshold, respectively.
The $2^{++}$ state is a bound state below the di-$J/\psi$ threshold. 
Studying the lineshape from the compact tetraquark picture
and comparing with those from the molecular picture 
can tell how much we have learnt from the experimental data and where 
we should go.

   \vspace{2cm}

{\it Acknowledgements}~~
Discussions with Qifang l\"u, Xianhui Zhong are acknowledged.
This work is partly supported by Guangdong Major Project of Basic and Applied Basic Research No.~2020B0301030008,
the National Natural Science Foundation of China with Grant No.~12035007,
Science and Technology Program of Guangzhou No.~2019050001,
Guangdong Provincial funding with Grant No.~2019QN01X172.
Q.W. is also supported by the NSFC and the Deutsche Forschungsgemeinschaft (DFG, German
Research Foundation) through the funds provided to the Sino-German Collaborative
Research Center TRR110 ``Symmetries and the Emergence of Structure in QCD"
(NSFC Grant No. 12070131001, DFG Project-ID 196253076-TRR 110).

\appendix
\newpage

\section{Several huge tables}
Several huge tables are collected in this appendix.

\begin{table*}[!htbp]
    \caption{\label{tab:full_heavy_tetra_tab}
        The Hamiltonian (the third column) in the bases listed in Table~\ref{tab:wave_func_tetra},
        predicted mass spectra (the forth and fifth columns) for the $cc\bar{c}\bar{c}$,
        $bb\bar{b}\bar{b}$ and $bc\bar{b}\bar{c}$ tetraquarks with various $J^{PC}$s
        as well as their corresponding eigenvectors (the last column).}
    \begin{tabular}{@{}cccccc@{}}
        \hline\hline
        $J^{PC}$ & Tetraquark         & $H$[MeV]   & Mass[MeV]  & Error[MeV]  & Eigenvector \\ \hline
        \multirow{3}*{$0^{++}$}
                 & $cc\bar{c}\bar{c}$ & \CoppH     & \CoppM     & \CoppMerr   & \CoppV      \\
                 & $bb\bar{b}\bar{b}$ & \BoppH     & \BoppM     & \BoppMerr   & \BoppV      \\
                 & $bc\bar{b}\bar{c}$ & \BCOppH    & \BCOppM    & \BCOppMerr  & \BCOppV     \\  \hline
        \multirow{3}*{$1^{+-}$}
                 & $cc\bar{c}\bar{c}$ & [6137.30]  & [6137.30]  & [0.25]      & [1]         \\
                 & $bb\bar{b}\bar{b}$ & [18889.81] & [18889.81] & [1.07]      & [1]         \\
                 & $bc\bar{b}\bar{c}$ & \BCIpmH    & \BCIpmM    & \BCIpmMerr  & \BCIpmV     \\  \hline
        $1^{++}$ & $bc\bar{b}\bar{c}$ & \BCIppH    & \BCIppM    & \BCIppMerr  & \BCIppV     \\  \hline
        \multirow{3}*{$2^{++}$}
                 & $cc\bar{c}\bar{c}$ & [6193.80]  & [6193.80]  & [0.35]      & [1]         \\
                 & $bb\bar{b}\bar{b}$ & [18920.61] & [18920.61] & [1.47]      & [1]         \\
                 & $bc\bar{b}\bar{c}$ & \BCIIppH   & \BCIIppM   & \BCIIppMerr & \BCIIppV
        \\  \hline\hline
    \end{tabular}
\end{table*}

\begin{table*}[!htbp]
    \caption{\label{tab:overlap}The overlaps between a tetraquark state and its possible decay channels}
    \begin{tabular}{@{}ccccclccc@{}}
        \hline\hline
        ~        & ~                                      & \multicolumn{3}{c}{$c\bar{c}\otimes c\bar{c}$} & ~                                              & \multicolumn{3}{c}{$b\bar{b}\otimes b\bar{b}$}                                                                                                           \\ \hline
        $J^{PC}$ & Tetraquark                             & $\eta_{c}\eta_{c}$                             & $J/\psi J/\psi$                                & $\eta_{c}J/\psi$                               & Tetraquark                             & $\eta_{b}\eta_{b}$ & $\Upsilon\Upsilon$ & $\eta_{b}\Upsilon$   \\\hline
        \multirow{2}*{$0^{++}$}
                 & $X_{cc\bar{c}\bar{c}}^{0^{++}}(6035)$  & 0.644                                          & ---                                            & ---                                            & $X_{bb\bar{b}\bar{b}}^{0^{++}}(18834)$ & 0.644              & ---                & ---                  \\
                 & $X_{cc\bar{c}\bar{c}}^{0^{++}}(6254)$  & 0.041                                          & -0.743                                         & ---                                            & $X_{bb\bar{b}\bar{b}}^{0^{++}}(18953)$ & 0.041              & -0.743             & ---
        \\\hline
        $1^{+-}$ & $X_{cc\bar{c}\bar{c}}^{1^{+-}}(6137)$  & ---                                            & ---                                            & 0.408                                          & $X_{bb\bar{b}\bar{b}}^{1^{+-}}(18890)$ & ---                & ---                & 0.408                \\\hline
        $2^{++}$ & $X_{cc\bar{c}\bar{c}}^{2^{++}}(6194)$  & ---                                            & 0.577                                          & ---                                            & $X_{bb\bar{b}\bar{b}}^{2^{++}}(18921)$ & ---                & 0.577              & ---                  \\\hline
        ~        & ~                                      & \multicolumn{4}{c}{$b\bar{b}\otimes c\bar{c}$} & \multicolumn{3}{c}{$b\bar{c}\otimes \bar{b}c$}                                                                                                                                                            \\\hline
        ~        & ~                                      & $\eta_{b}\eta_{c}$                             & $\eta_{b}J/\psi$                               & $\Upsilon\eta_{c}$                             & $\Upsilon J/\psi$                      & $B_{c}B_{c}$       & $B_{c}^{*}B_{c}$   & $B_{c}^{*}B_{c}^{*}$ \\
        \multirow{4}*{$0^{++}$}
                 & $X_{bc\bar{b}\bar{c}}^{0^{++}}(12516)$ & 0.641                                          & ---                                            & ---                                            & ---                                    & ---                & ---                & ---                  \\
                 & $X_{bc\bar{b}\bar{c}}^{0^{++}}(12646)$ & 0.072                                          & ---                                            & ---                                            & ---                                    & -0.072             & ---                & ---                  \\
                 & $X_{bc\bar{b}\bar{c}}^{0^{++}}(12431)$ & 0.762                                          & ---                                            & ---                                            & ---                                    & ---                & ---                & ---                  \\
                 & $X_{bc\bar{b}\bar{c}}^{0^{++}}(12595)$ & -0.0546                                        & ---                                            & ---                                            & ---                                    & -0.0546            & ---                & ---                  \\\hline
        \multirow{4}*{$1^{+-}$}
                 & $X_{bc\bar{b}\bar{c}}^{1^{+-}}(12363)$ & ---                                            & ---                                            & ---                                            & ---                                    & ---                & ---                & ---                  \\
                 & $X_{bc\bar{b}\bar{c}}^{1^{+-}}(12597)$ & ---                                            & -0.308                                         & 0.308                                          & ---                                    & ---                & ---                & ---                  \\
                 & $X_{bc\bar{b}\bar{c}}^{1^{+-}}(12339)$ & ---                                            & ---                                            & ---                                            & ---                                    & ---                & ---                & ---                  \\
                 & $X_{bc\bar{b}\bar{c}}^{1^{+-}}(12573)$ & ---                                            & 0.408                                          & 0.408                                          & ---                                    & ---                & ---                & ---                  \\\hline
        \multirow{2}*{$1^{++}$}
                 & $X_{bc\bar{b}\bar{c}}^{1^{++}}(12538)$ & ---                                            & ---                                            & ---                                            & ---                                    & ---                & ---                & ---                  \\
                 & $X_{bc\bar{b}\bar{c}}^{1^{++}}(12618)$ & ---                                            & ---                                            & ---                                            & 0.947                                  & ---                & ---                & ---                  \\\hline
        \multirow{2}*{$2^{++}$}
                 & $X_{bc\bar{b}\bar{c}}^{1^{++}}(12597)$ & ---                                            & ---                                            & ---                                            & 0.817                                  & ---                & ---                & ---                  \\
                 & $X_{bc\bar{b}\bar{c}}^{1^{++}}(12613)$ & ---                                            & ---                                            & ---                                            & 0.477                                  & ---                & ---                & ---
        \\\hline\hline
    \end{tabular}
\end{table*}

\begin{table*}[!htbp]
    \caption{\label{tab:decay_width_tetra}The values of $\frac{1}{(2S+1)8\pi}|\mathcal{M}_{i}^{(j)}|^{2}\frac{|\bm{p}|}{M^{2}}$ for the $cc\anticc$, $bb\antibb$, $bc\antibc$ tetraquark states}
    \begin{tabular}{@{}ccccclccc@{}}
        \hline\hline
        ~        & ~                                      & \multicolumn{3}{c}{$c\bar{c}\otimes c\bar{c}$} & ~                                              & \multicolumn{3}{c}{$b\bar{b}\otimes b\bar{b}$}                                                                                                           \\ \hline
        $J^{PC}$ & Tetraquark                             & $\eta_{c}\eta_{c}$                             & $J/\psi J/\psi$                                & $\eta_{c}J/\psi$                               & Tetraquark                             & $\eta_{b}\eta_{b}$ & $\Upsilon\Upsilon$ & $\eta_{b}\Upsilon$   \\\hline
        \multirow{2}*{$0^{++}$}
                 & $X_{cc\bar{c}\bar{c}}^{0^{++}}(6035)$  & \num{2.03e-4}                                  & ---                                            & ---                                            & $X_{bb\bar{b}\bar{b}}^{0^{++}}(18834)$ & \num{2.726e-5}     & ---                & ---                  \\
                 & $X_{cc\bar{c}\bar{c}}^{0^{++}}(6254)$  & \num{1.599e-6}                                 & \num{2.431e-4}                                 & ---                                            & $X_{bb\bar{b}\bar{b}}^{0^{++}}(18953)$ & \num{2.259e-7}     & \num{3.409e-5}     & ---
        \\\hline
        $1^{+-}$ & $X_{cc\bar{c}\bar{c}}^{1^{+-}}(6137)$  & ---                                            & ---                                            & \num{2.435e-5}                                 & $X_{bb\bar{b}\bar{b}}^{1^{+-}}(18890)$ & ---                & ---                & \num{3.336e-6}       \\\hline
        $2^{++}$ & $X_{cc\bar{c}\bar{c}}^{2^{++}}(6194)$  & ---                                            & ---                                            & ---                                            & $X_{bb\bar{b}\bar{b}}^{2^{++}}(18921)$ & ---                & \num{7.198e-8}     & ---                  \\\hline
        ~        & ~                                      & \multicolumn{4}{c}{$b\bar{b}\otimes c\bar{c}$} & \multicolumn{3}{c}{$b\bar{c}\otimes \bar{b}c$}                                                                                                                                                            \\\hline
        ~        & ~                                      & $\eta_{b}\eta_{c}$                             & $\eta_{b}J/\psi$                               & $\Upsilon\eta_{c}$                             & $\Upsilon J/\psi$                      & $B_{c}B_{c}$       & $B_{c}^{*}B_{c}$   & $B_{c}^{*}B_{c}^{*}$ \\
        \multirow{4}*{$0^{++}$}
                 & $X_{bc\bar{b}\bar{c}}^{0^{++}}(12516)$ & \num{2.049e-3}                                 & ---                                            & ---                                            & ---                                    & ---                & ---                & ---                  \\
                 & $X_{bc\bar{b}\bar{c}}^{0^{++}}(12646)$ & \num{3.589e-5}                                 & ---                                            & ---                                            & ---                                    & \num{2.5e-5}       & ---                & ---                  \\
                 & $X_{bc\bar{b}\bar{c}}^{0^{++}}(12431)$ & \num{1.76e-3}                                  & ---                                            & ---                                            & ---                                    & ---                & ---                & ---                  \\
                 & $X_{bc\bar{b}\bar{c}}^{0^{++}}(12595)$ & \num{1.864e-5}                                 & ---                                            & ---                                            & ---                                    & \num{9.805e-6}     & ---                & ---                  \\\hline
        \multirow{4}*{$1^{+-}$}
                 & $X_{bc\bar{b}\bar{c}}^{1^{+-}}(12363)$ & ---                                            & ---                                            & ---                                            & ---                                    & ---                & ---                & ---                  \\
                 & $X_{bc\bar{b}\bar{c}}^{1^{+-}}(12597)$ & ---                                            & \num{5.47e-6}                                  & \num{6.647e-6}                                 & ---                                    & ---                & ---                & ---                  \\
                 & $X_{bc\bar{b}\bar{c}}^{1^{+-}}(12339)$ & ---                                            & ---                                            & ---                                            & ---                                    & ---                & ---                & ---                  \\
                 & $X_{bc\bar{b}\bar{c}}^{1^{+-}}(12573)$ & ---                                            & \num{8.401e-6}                                 & \num{1.073e-5}                                 & ---                                    & ---                & ---                & ---                  \\\hline
        \multirow{2}*{$1^{++}$}
                 & $X_{bc\bar{b}\bar{c}}^{1^{++}}(12538)$ & ---                                            & ---                                            & ---                                            & ---                                    & ---                & ---                & ---                  \\
                 & $X_{bc\bar{b}\bar{c}}^{1^{++}}(12618)$ & ---                                            & ---                                            & ---                                            & \num{3.987e-5}                         & ---                & ---                & ---                  \\\hline
        \multirow{2}*{$2^{++}$}
                 & $X_{bc\bar{b}\bar{c}}^{1^{++}}(12597)$ & ---                                            & ---                                            & ---                                            & \num{1.436e-5}                         & ---                & ---                & ---                  \\
                 & $X_{bc\bar{b}\bar{c}}^{1^{++}}(12613)$ & ---                                            & ---                                            & ---                                            & \num{5.813e-6}                         & ---                & ---                & ---
        \\\hline\hline
    \end{tabular}
\end{table*}


\end{document}